\documentclass[10pt]{article}
\usepackage{amsmath}
\usepackage{amssymb} 
\usepackage{caption}
\usepackage{graphicx}
\usepackage{authblk}
\usepackage[hypertexnames=false,colorlinks=true,linkcolor=blue,citecolor=blue]{hyperref}
\usepackage[numbers,comma,square,sort&compress]{natbib}
\usepackage[letterpaper,text={6.5in,9in},centering]{geometry}

\graphicspath{{eps/}{pdf/}}

\makeatletter\@addtoreset{equation}{section}\makeatother

\newtheorem{Lemma}{Lemma}[section]

\newtheorem{Proposition}[Lemma]{Proposition}
\newtheorem{Corollary}[Lemma]{Corollary}
\newtheorem{Remark}[Lemma]{Remark}

\newtheorem{Hypothesis}[Lemma]{Hypothesis}

\newenvironment{Proof}%
 {\begin{trivlist} \item[]{\bf Proof. }}%
 {\hspace*{\fill}$\rule{.4\baselineskip}{.4\baselineskip}$\end{trivlist}}

\newenvironment{Acknowledgment}%
 {\begin{trivlist}\item[]\textbf{Acknowledgments.}}{\end{trivlist}}
\newcommand{\R}{\mathbb{R}}

\newcommand{\N}{\mathbb{N}}
\newcommand{\Z}{\mathbb{Z}}

\def\Re{\mathop{\mathrm{Re}}}
\def\Im{\mathop{\mathrm{Im}}}

\newcommand{\rmO}{\mathrm{O}}

\newcommand{\rmd}{\mathrm{d}}
\newcommand{\rme}{\mathrm{e}}
\newcommand{\rmi}{\mathrm{i}}

\newcommand{\Rg}{\mathrm{Rg}}

\usepackage{tikz}

\begin{document}

\title{Continuation and Bifurcation of Grain Boundaries in the Swift-Hohenberg Equation}
\author[1]{David J.B. Lloyd}
\author[2]{Arnd Scheel}
\affil[1]{\small Department of Mathematics, University of Surrey, Guildford, GU2 7XH, UK}
\affil[2]{\small School of Mathematics, University of Minnesota, 206 Church Street S.E, Minneapolis, MN 55455}
\date{\today}
\maketitle

\begin{abstract}
\noindent
We study grain boundaries between striped phases in the prototypical Swift-Hohenberg equation. We propose an analytical and numerical far-field-core decomposition that allows us to study existence and bifurcations of grain boundaries analytically and numerically using continuation techniques. This decomposition overcomes problems with computing grain boundaries in a large doubly periodic box with phase conditions. Using the spatially conserved quantities of the time-independent Swift-Hohenberg equation, we show that symmetric grain boundaries must select the marginally zig-zag stable stripes. We find that as the angle between the stripes is decreased, the symmetric grain boundary undergoes a parity-breaking pitchfork bifurcation where dislocations at the grain boundary split into disclination pairs. A plethora of asymmetric grain boundaries (with different angles of the far-field stripes either side of the boundary) is found and investigated. The energy of the grain boundaries is then mapped out. We find that when the angle between the stripes is greater than a critical angle, the symmetric grain boundary is energetically preferred while when the angle is less than the critical angle, the grain boundaries where stripes on one side are parallel to the interface are energetically preferred. Finally, we propose a classification of grain boundaries that allows us to predict various non-standard asymmetric grain boundaries.

\end{abstract}

\section{Introduction}

Grain boundaries are a basic building block for spatial patterns in extended systems; see for instance~\cite{manneville1990,cross1993,hoyle2006}. They separate regions in physical space, where the fine crystalline structure possesses different orientations. While they are extensively studied in many aspects of material science, they also arise in pattern-forming systems such as Rayleigh-B\'enard convection. Our focus here is on the latter, pattern-forming systems far from thermodynamic equilibrium, although we suspect that many of the methods here can be applied to crystalline patterns in interacting-particle systems, say. 

\begin{figure}[h]
	\centering
	\includegraphics[width=0.7\linewidth]{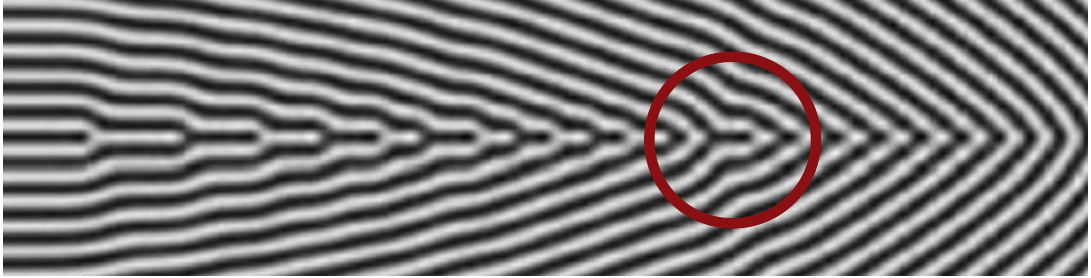}
	\caption{Zoom-in of Figure 8 from Ercolani {\it et al}~\cite{ercolani2003}, which shows a simulation of the Swift-Hohenberg equation \eqref{e:sh}, $\mu=1$,  in an ellipsoidal domain after initial transient. One notices the qualitative change of grain boundaries along the horizontal axis, highlighted by the red circle, as the angle between stripes becomes more acute; see Section \ref{s:path} for more details.}\label{f:stadion}
\end{figure}

Our motivation is two-fold. First, coherent structures in systems far from equilibrium have been studied quite successfully recently using a spatial-dynamics point of view; see for instance \cite{sandstede2004}. These methods have proven useful not only to establish local existence, but also to classify and study stability, bifurcations, and interactions of coherent structures. This spatial-dynamics perspective has also been  used to study existence of grain boundaries close to onset of a pattern-forming instability \cite{haragus2007,haragus2012,scheel2014}. Second, far from onset of pattern formation, qualitative changes in the nature of grain boundaries have been observed and quantified, both theoretically and numerically in \cite{passot1994,ercolani2003,ercolani2009} using phase approximations. Figure \ref{f:stadion} \cite{ercolani2003} shows a direct simulation of the Swift-Hohenberg equation in an ellipsoidal domain, with boundary conditions forcing parallel stripes. Along the major axis, weak bending of stripes is eventually mediated by grain boundaries and defects. As curvature and hence angles of grain boundaries increases inwards, the grain boundaries go through a sequence of qualitative changes that motivated the studies mentioned above and our computations here. 

The main purpose of this paper is also twofold. First, we lay out a systematic numerical strategy for the study of grain boundaries, inspired very much by the spatial dynamics point of view where grain boundaries are heteroclinic orbits. Second, we study grain boundaries in the prototypical example of the Swift-Hohenberg equation numerically. Our approach is based on numerical continuation with ``asymptotic boundary conditions'', at zeroth order. It enables us to cleanly separate the core of the grain boundary from far-field behavior, and thereby allows us to detect bifurcations  in a ``thermodynamic limit'' of infinite domain size. More practically, it allows us to construct a well-posed continuation problem with well-conditioned linear operators, uniformly in the domain size. One of our most striking observations concerns the behavior of grain boundaries as the angle between the stripes is decreased towards an acute angle.  Decreasing the angle as a continuation parameter, we first locate a parity-breaking super-critical pitchfork bifurcation. The asymmetric branch breaks a parity-shift symmetry and quickly develops into a convex-concave disclination pair. The primary branch later develops two dislocations, and restabilizes shortly after.

The remainder of the introduction recalls basic facts about the Swift-Hohenberg equation, previous results about existence of grain boundaries and defects, and gives a brief outline of the paper.

\paragraph{The Swift-Hohenberg model.}
We study grain boundaries in the Swift-Hohenberg equation,
\begin{equation}\label{e:sh}
u_t=-(\Delta +1)^2 u + \mu u - u^3, \qquad (x,y)\in\R^2,
\end{equation}
as a prototypical model for the formation of striped phases. The trivial state, $u(x,y)\equiv 0$ is linearly unstable against perturbations of the form $\rme^{\rmi (k_x x + k_y y)}$, for $k_x^2+k_y^2\sim 1$, $\mu\gtrsim 0$. Stable solutions in this regime are striped (or roll) solutions $u_\mathrm{s}(kx;k)$, $u_\mathrm{s}(\xi)=u_\mathrm{s}(\xi+2\pi)=u_\mathrm{s}(-\xi)$, which exist for an interval of allowed wavenumbers $k\in (k_\mathrm{min},k_\mathrm{max})$ and are stable for $k\in (k_\mathrm{zz},k_\mathrm{eck})$. Here, $k_\mathrm{eck}=k_\mathrm{eck}(\mu)$ and $k_\mathrm{zz}=k_\mathrm{zz}(\mu)$ denote Eckhaus (instability due to perturbations of wavelength) and zigzag (instability due to transverse perturbations) boundaries, respectively, with leading order expansions
\[
k_\mathrm{min,max}=1\pm\sqrt{\mu/4},\quad  k_\mathrm{eck}=1+\sqrt{\mu/12},\quad k_\mathrm{zz}=1-\mu^2/512;
\]
see for instance \cite{cross1993,mielke1997}. 

While individual striped solutions are asymptotically stable \cite{schneider1996,uecker1999}, one typically observes patches of stripe solutions with different orientation in a large domain. Indeed, any rotated stripe solution $u_\mathrm{s}(k_x x + k_y y;k)$, $k_x^2+k_y^2=k^2$, is a solution due to rotational symmetry of \eqref{e:sh}. We are interested in situations where two different orientations $\underline{k}^\pm=(k_x^\pm,k_y^\pm)$ are dominant in, say, $x>0$ and $x<0$, respectively, separated by an exponentially localized interfacial region near $x\sim 0$, which we will refer to as a \emph{grain boundary}, thinking of the orientation of stripes as the grain  or microstructure in the medium. 

Grain boundaries often possess a vertical periodicity. In particular, when $k_y^\pm$ are commensurate, $k_y^-/q_-=k_y^+/q_+$ for some integer $q_\pm$, then stripes at $\pm\infty$ possess a common periodicity $L_y=2\pi/k_y$, $k_y=k_y^\pm/q_\pm$. In this case, one can view grain boundaries as heteroclinic orbits to asymptotic periodic orbits,
\begin{equation}\label{e:gbdef}
\left|u_\mathrm{gb}(x+\xi,y)-u_\mathrm{s}(k_y^\pm y + k_x^\pm x+\xi+\varphi^\pm;k^\pm)\right|_{X_\mathrm{loc}}\to 0,\quad \xi\to \pm\infty,
\end{equation}
where norms could be taken in $X_\mathrm{loc}=H^4([0,1]\times S^1)$, $S^1=\R/2\pi\Z$, in the independent variables $x,y$, and $(k^\pm)^2=(k^\pm_x)^2+(k^\pm_y)^2$. 

\begin{figure}[h]
\centering
\includegraphics[width=.8\linewidth]{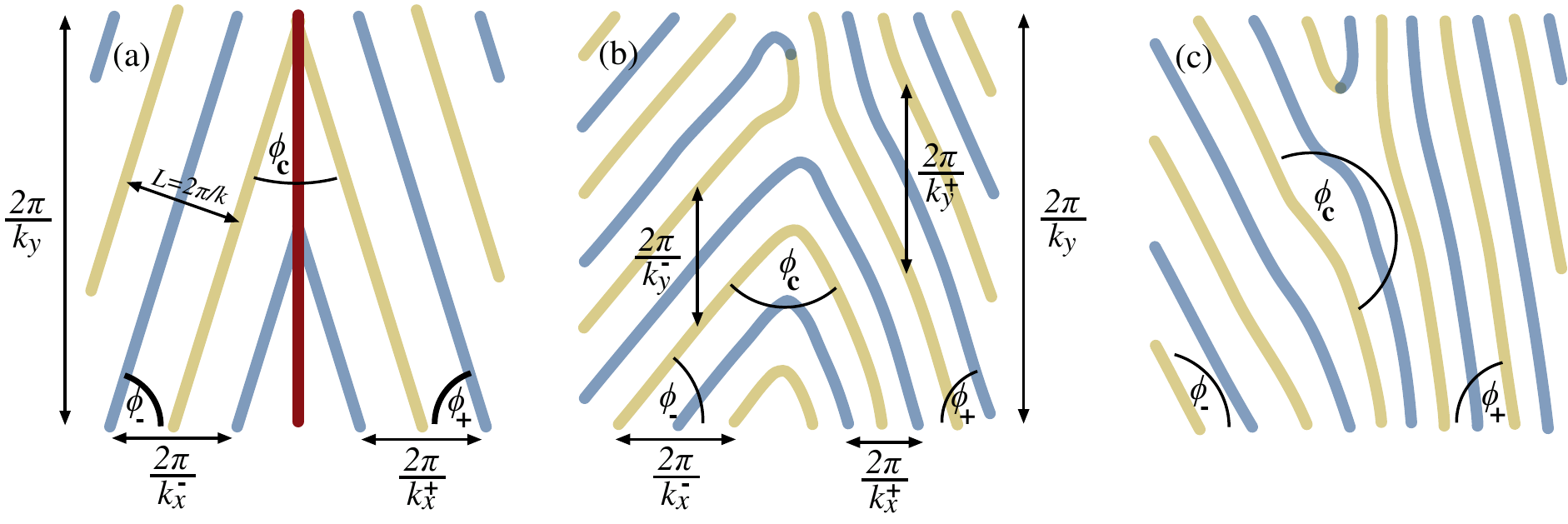}
\caption{Schematic figures of grain boundaries, when  $(q_-,q_+)$ is  (a) $(1,-1)$ (b) $(3,-2)$, (c) $(2,1)$. Note that the $q_\pm$ count the number of stripes encountered in a fixed section $x=\pm L$, $L$ large, $y\in (0,L_y)$.}
\label{f:gbsc}
\end{figure}

We refer to such a solution and associated $q_\pm$ as a \emph{$(q_-,q_+)$ grain boundary}; see Figure \ref{f:gbsc}. We also use a convention where the sign of $q_\pm$ indicates positive and negative slope of level sets as graphs over $x$, respectively. Since we can reflect vertically, in $y$, we adopt the convention where $q_->0$.

\paragraph{Small-amplitude grain boundaries: Normal forms.}
Intuitively, it is not immediately clear that time-independent equilibria of the form  \eqref{e:gbdef} actually exist for say the Swift-Hohenberg equation in an idealized unbounded domain. One could easily envision how the curvature along a family of stripes decreases slowly in time until stripes are straight. 

Mathematically, the question of existence was answered quite comprehensively in \cite{haragus2007,haragus2012,scheel2014}. There, existence of symmetric grain boundaries, $k_x^-=-k_x^+$, $k_y^-=k_y^+$ was shown for $\mu$ sufficiently small and arbitrary angle $\angle(\underline{k}^-,\underline{k}^+)$. The approach there reformulates the stationary Swift-Hohenberg equation in a strip
\begin{equation}\label{e:shgb}
-(\Delta +1)^2 u + \mu u - u^3=0, \qquad x\in\R,\ \ y\in\R/\left(\frac{2\pi}{k_y}\right)\Z,
\end{equation}
as an (ill-posed) dynamical system in the $x$-direction, formally writing it as a first-order equation in $x$, 
\begin{equation}\label{e:ds}
\frac{dU}{dx} = \mathcal A(\mu,k)U + \mathcal F(U),
\end{equation}
in which
\[
U = \begin{pmatrix}u\\u_1\\v\\v_1\end{pmatrix},\quad
\mathcal A(\mu,k) = \begin{pmatrix}
0&1&0&0\\ -(1+k_y^2\partial_y^2)&0&1&0\\ 0&0&0&1\\
\mu& 0&
-(1+k_y^2\partial_y^2)&0\end{pmatrix},\quad 
\mathcal F(U) = \begin{pmatrix}0\\0\\0\\-u^3\end{pmatrix}.
\]
Here $U$ takes values in Sobolev spaces of periodic functions $U\in \prod_{j=0}^3 H^{3-j}_\mathrm{per}(0,2\pi)$
and $y$ was rescaled with $k_y$ to be of period $2\pi$. Grain boundaries are now heteroclinic orbits in the traditional dynamical systems sense, where the (infinite-dimensional) phase-space variable $U(x)$ converges to periodic orbits $U_\mathrm{r}^\pm$ for $x\to\pm\infty$. 

The results in \cite{haragus2007,haragus2012,scheel2014} examine this dynamical system using center-manifold reduction and normal form theory. The ill-posed dynamical systems \eqref{e:ds} is reduced to an ordinary differential equation on a locally invariant manifold. The dynamics on this center-manifold describe the spatial ($x$-)evolution of profiles $U(x,y)$. In order to analyze these dynamics, normal form coordinate changes, analogous to averaging theory are employed, which eventually exhibit invariant subspaces within a higher-dimensional system of differential equations. In normal form, the reduced equation consists of  coupled, stationary Ginzburg-Landau equations, which capture amplitudes of modes $\rme^{\rmi (\kappa_x x+\kappa_y y)}$, where $\kappa_x^2+\kappa_y^2=1$, and  $\kappa_y=\ell k_y$, $\ell \in \Z$. Invariant subspaces amount to setting amplitudes associated with $\ell\neq 1$ to zero and restricting to real amplitudes. The normal form equations had been derived much earlier, starting with the assumption that relevant modes consist \emph{only} of two differently oriented stripes, $\ell=\pm 1$, whose dynamics is then well described by a Newell-Whitehead-Segel amplitude equation \cite{malomed1990}.

After suitable scalings, the normal form equations read (assuming a non-resonance condition on the angle,  $1/k_y\not\in\Z$),
\begin{equation}\label{e:nf}
\kappa_\ell^2 (C_\ell)''=-C_\ell(1-2\sum_{\ell'\neq \ell,\pm}|C_{\ell'}|^2-|C_{\ell}|^2),\quad  |\ell|<1/k_y 
\end{equation}
where $\kappa_\ell=\mathrm{sign}(\ell)\sqrt{1-\ell^2 k_y^2}$, $\ell\in \Z$; see \cite{scheel2014} for details. 

These normal form equations possess pure mode equilibria $\underline{C}^*$ with $C_\ell=1$ for $\ell=\ell_*$, $C_\ell=0$ otherwise, which simply correspond to slanted stripes $\rme^{\rmi (k_x x+\ell_* k_y y)}$ with $\ell_*$ maxima of $u$ across any section $x=x_0$, $y\in (0,2\pi)$. More interestingly, they also possess heteroclinic orbits connecting any two pure-more equilibria $\underline{C}^+$ and $\underline{C}^-$ with $\ell_*=\ell^\pm$; \cite{scheel2014,vandenberg2000,weth2013}. Asymptotic states of these heteroclinics at $x=\pm\infty$ possesses different orientations relative to the grain boundary and correspond to $(\ell_-,\ell_+)$ grain boundaries in our terminology;  see Figure \ref{f:gbsc}.

\paragraph{Wavenumber selection.}

\begin{figure}[h]
	\centering
	\includegraphics[width=1.0\linewidth]{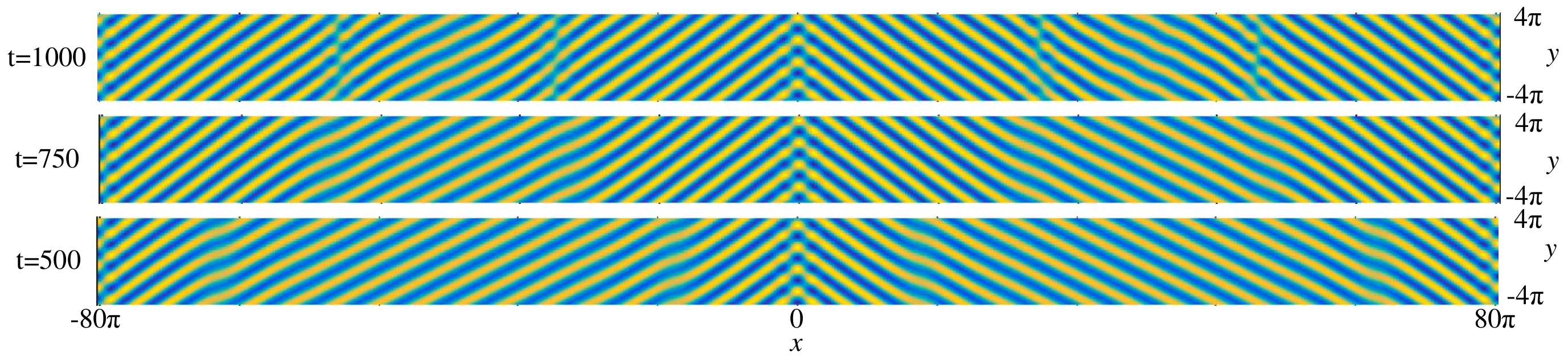}
	\caption{Snap shots of the time evolution of \eqref{e:sh} with $\mu=0.1$ in a doubly periodic box  $[-80\pi,80\pi)\times[0,\pi/k_y)$, $k_y=0.8$, $N_x=2^{10},N_y=2^6$. Initial far-field stripes have an asymptotic wavenumber of 0.9. Grain boundaries select $k\sim 1$. Note that $x$-periodicity enforces two grain boundaries and the corrected wavenumber spreads from both into the bulk.  }\label{f:sn}
\end{figure}

Inspecting the heteroclinics, in the leading-order amplitude equation, one finds that heteroclinics connect equilibria, only, not nearby periodic orbits $C_{\ell_*}\sim \rme^{\rmi \varepsilon x}$. In other words, grain boundaries select wavenumbers in the far field. The stripes are well described by a nonlinear phase-diffusion equation far from the grain boundary. The effect of the grain boundary can then be thought of as an inhomogeneous Neumann boundary condition for the phase or, equivalently, an inhomogeneous Dirichlet boundary condition for the wavenumber. The effect of this inhomogeneous boundary condition spreads diffusively through the domain. This effect was illustrated in an amplitude approximation in \cite{malomed1990}. We demonstrate the diffusive spread in the Swift-Hohenberg equation in Figure \ref{f:sn}. We initialize the system in a strip with wavenumber $k=0.9$ away from the grain boundary. One clearly sees a change in wavenumber spreading from the grain boundaries into the domain, causing intermittent phase slips. 
 
\begin{figure}[h]
	\centering
	\includegraphics[width=0.4\linewidth]{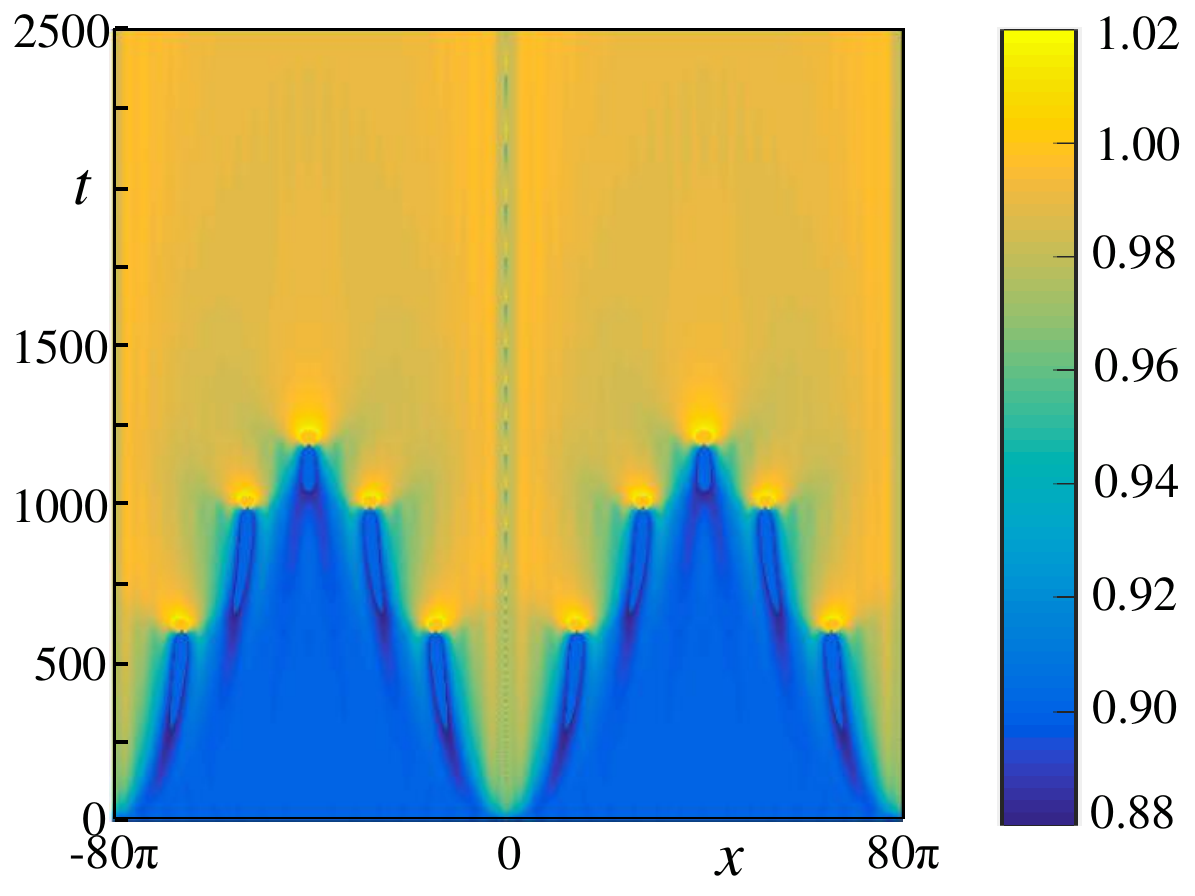}
	\caption{Space-time contour plot of the instantaneous wavenumber of the solution from Figure \ref{f:sn}. One clearly sees a diffusive spread of the selected wavenumber $k\sim 1$ from the two grain boundaries, with the intermittent phase slips as singularities of the wavenumber at approximately $t\sim 602,993,1192$.}\label{f:cont}
\end{figure}

We extract the change in wavenumber from the solution directly, by computing first the analytic signal for fixed $y=y_0$, $z(x,y_0)=u(x,y_0)+\rmi \mathcal{H} u(\cdot,y_0)(x)$, where $\mathcal{H}$ is the Hilbert transform, and then extracting the wavenumber as $k(x,y_0)=(\Im\log z)'(x,y_0)$. We finally average over $y_0$ to obtain $\bar{k}(x)$ at each time step. A contour space-time plot of $\bar{k}$ is shown in Figure \ref{f:cont}. 

\paragraph{Symmetries and phase matching.}
Before investigating multiplicities of grain boundaries more closely, we recall the underlying relevant symmetries. The Swift-Hohenberg equation is invariant under translations $T^x_\xi$ and $T_\xi^y$ and reflections $R^x$ and $R^y$ in $x$ and $y$, respectively, and also possesses the up-down, or parity symmetry $S:u\mapsto -u$. Grain boundaries therefore necessarily come in two-parameter families, induced by translation in $x$ and $y$. In the center-manifold reduced equations, $y$-translations act as complex rotations on $C_\ell$, $y$-reflection conjugates $C_{\ell}$ and $C_{-\ell}$. An additional normal form symmetry allows independent complex rotations in all amplitudes $C_\ell$ at leading order. As a consequence, grain boundaries come in a degenerate 3-parameter family, where one can arbitrarily shift stripes on either side of the grain boundary parallel to the boundary. One expects that terms beyond the normal form would yield conditions for this relative shift. In \cite{haragus2007,haragus2012,scheel2014}, the reflection symmetry was used to show that grain boundaries that are symmetric in $x$ persist for the full system. 

\begin{figure}[h]
\centering
\includegraphics[width=0.7\linewidth]{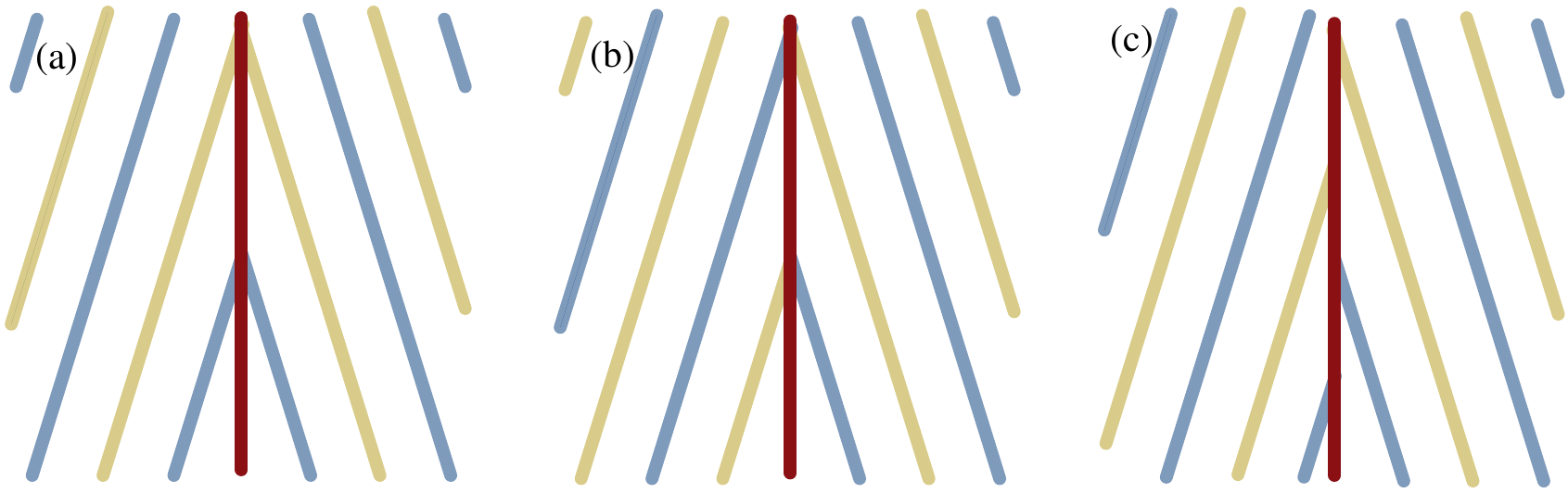}
\caption{Schematic sketch of phase matching, (a) in-phase (b) anti-phase, and (c) phase mismatch. Solutions (a) and (b) were proven to exist in \cite{haragus2012,scheel2014} and are computed here, existence of mismatched solutions is not known. }
\label{f:gbph}
\end{figure}

Using the same methods, one can also show that grain boundaries that are invariant under $R^xS$, $x\mapsto -x$, $u\mapsto -u$ exist in the full equation. We refer to these two types of grain boundaries as phase matched or anti-phase matched. It seems difficult to determine whether other $(1,-1)$ grain boundaries exist at small amplitudes; see Figure \ref{f:gbph} for an illustration of phase matched and anti-phase matched grain boundaries. 

On the other hand,  one can find asymmetric $(q_-,q_+)$ grain boundaries, connecting $C_{q_-}$ and $C_{q_+}$ in the normal form,  using variational methods (see \cite{weth2013}; the existence is also stated in \cite{malomed1990}). Existence for the full equation then relies on solving a phase matching equation for the relative shift of stripes at the interface. Our numerical results below strongly indicate that such solutions do actually persist, that is, one can solve the phase-matching equation. 

We emphasize that the results in \cite{scheel2014} yield existence of grain boundaries for arbitrary $k_y$ (effectively changing the angle), and $\mu<\mu_*(k_y)$ sufficiently small. However, since  $\mu_*(k_y)\to 0$ for $k_y\to 0$, the results do not imply that grain boundaries exist for arbitrary angle and fixed $\mu>0$, sufficiently small.

\paragraph{Grain boundaries: Bifurcations.} 

\begin{figure}[h]
\centering
\includegraphics[width=0.7\linewidth]{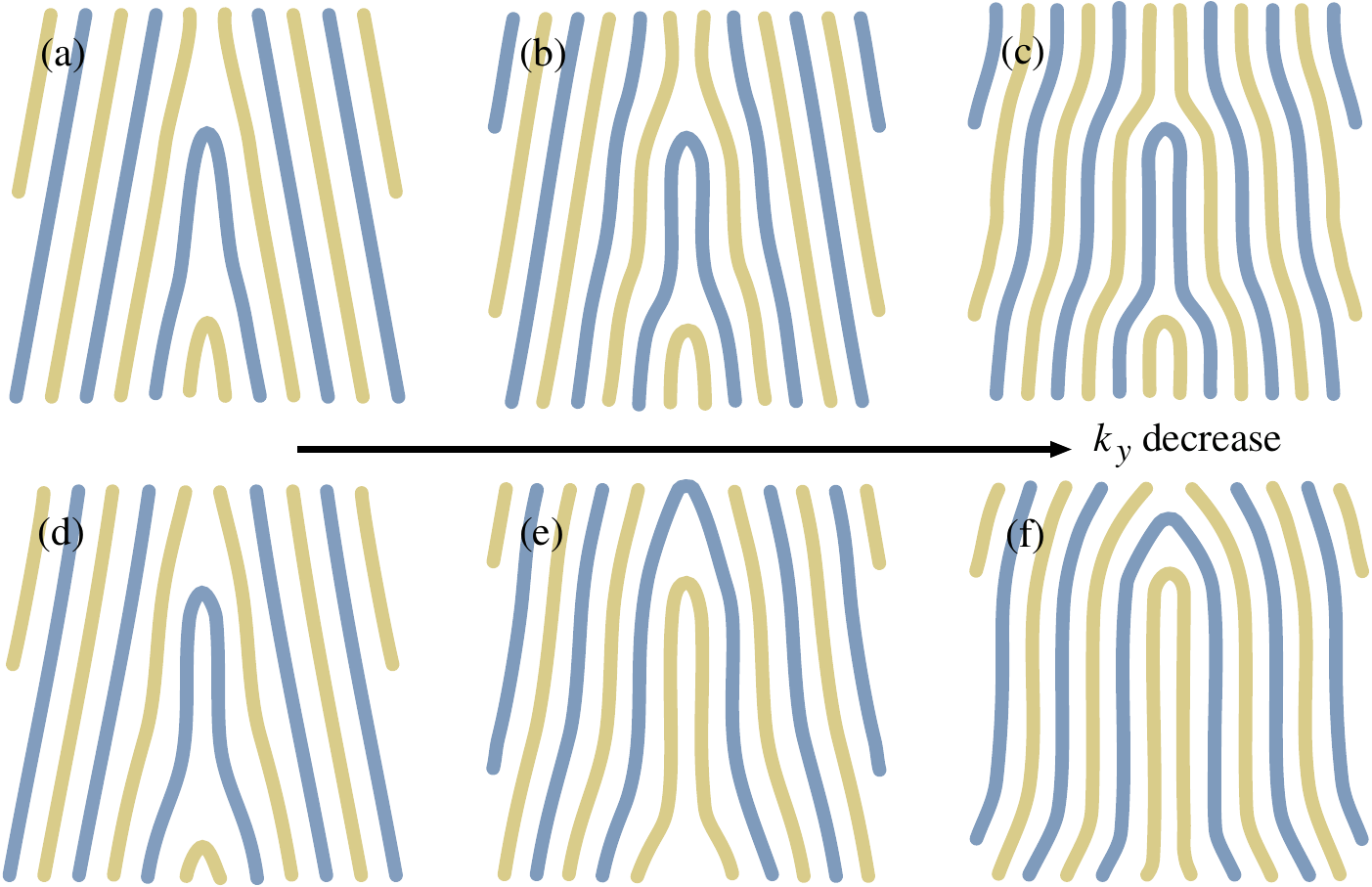}
\caption{Schematic picture of  the formation of protrusions and defects at the grain boundary.}\label{f:acute}
\end{figure}

Our present study is motivated to a large extent by work on grain boundaries at finite amplitude, which predicts intriguing qualitative changes as the angle between the stripes becomes more acute; see the grain boundaries circled in red in Figure \ref{f:stadion} and the schematics in Figure \ref{f:acute}. In the weak bending regime, it has been well known \cite{cross1993,haragus2007} that grain boundaries can be described within the Cross-Newell phase approximation. 
In other words, the change in orientation is well described to leading order by a slow change in orientation of the stripes across the interface. For more acute angles, a phase transition occurs, when protrusions form at the grain boundaries, effectively creating dislocations \cite{passot1994,ercolani2003,ercolani2009} or disclination pairs. The analysis in \cite{passot1994,ercolani2003,ercolani2009} predicts the onset of defect formation at the grain boundary theoretically with good accuracy, but is largely based on phase approximations which may loose validity near defects. On the other hand, the existence results in \cite{scheel2014} do not predict any bifurcations of grain boundaries. 

\begin{figure}[h]
\centering
\includegraphics[width=0.7\linewidth]{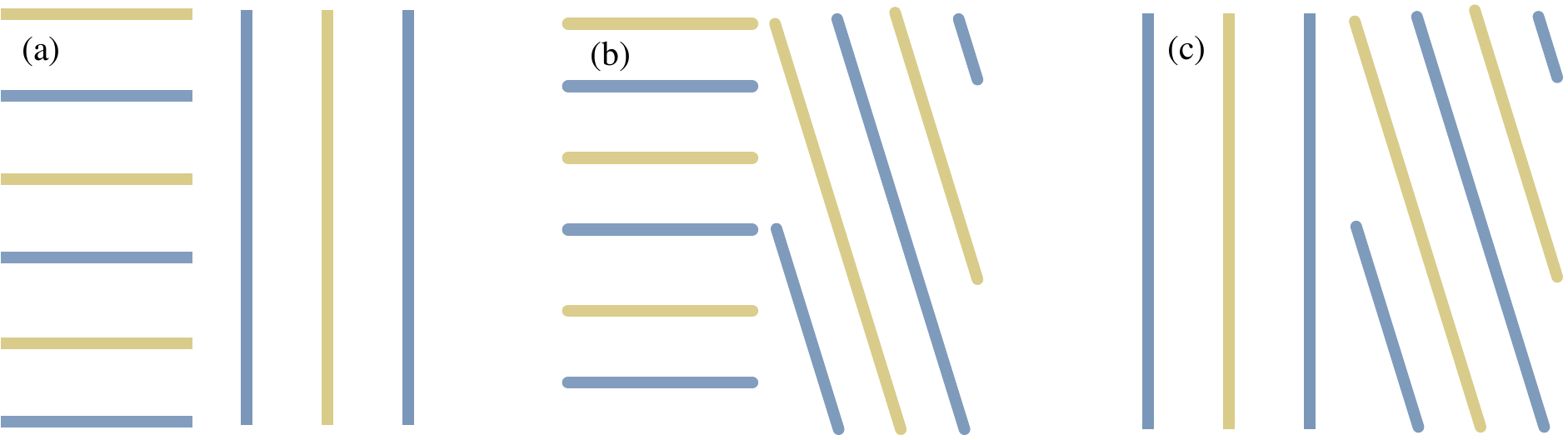}
\caption{Sketch of interesting limiting grain orientations, where stripes are oriented parallel of perpendicular to the grain boundary. }\label{f:perp}
\end{figure}

Furthermore, we study a wealth of asymmetric grain boundaries, attempting a systematic description in terms of defects, resonances, and pinning effects. We encounter interesting bifurcations near limiting cases, when grain boundaries are parallel or perpendicular to the orientation of one of the stripes; see Figure \ref{f:perp}. In fact, normal form equations are more difficult in these resonant cases \cite{scheel2014} and had been analyzed in \cite{manneville1983grain}. The dynamics near such grain boundaries are quite intricate, governed by non-adiabatic pinning effects \cite{vinals2007}. We present some numerical results near these configurations in Section \ref{s:hor}.

\paragraph{Numerical approaches.} There appear to be few systematic numerical studies of grain boundaries beyond direct simulations. Most detailed results were obtained in \cite{ercolani2003}, where grain boundaries along $x=0$ were computed by imposing oblique boundary conditions at $x=\pm L_x$ and by suppressing patterns for large $|y|$ via a parameter ramp. On the other hand, solutions to the leading-order normal form equations \eqref{e:nf} can be readily computed as heteroclinic orbits to an ordinary differential equation \cite{malomed1990,hargus2012a}. Implementing such a view point for \eqref{e:shgb}, one would project \eqref{e:shgb} onto functions $u_N(x,y)=\sum_{|\ell|\leq N} u_\ell(x)\rme^{\rmi \ell y}$. In the resulting system of ordinary differential equations, one would look for heteroclinic orbits that connect periodic orbits at $x=\pm\infty$. Pursuing this point of view, one would like to build on recent results in the dynamical systems literature on computation of heteroclinic orbits connecting periodic orbits and equilibria \cite{beyn1990,pampel2001,dieci2004,krauskopf2007,doedel2008a,doedel2008b}. Main ingredients in these approaches are a truncation to a finite interval $x\in (-L_x,L_x)$ with appropriate boundary conditions at $x=\pm L_x$, phase conditions that rule out translation symmetry in $x$ and other potential multiplicities, and finally appropriate discretizations of the ODE. 

For boundary conditions at $x=\pm L_x$, one wishes to require that the solution lies in the stable and unstable manifold of the asymptotic state, respectively. These local stable and unstable manifolds can then be approximated to first order by their tangent spaces. In the case of periodic orbits, this involves a somewhat cumbersome construction of Floquet bundles, in addition to actually computing the limiting periodic orbit. One can easily envision that such computations become tedious and slow when the dimension of the system $N$ tends to infinity. 

Our approach is similar in spirit, although it does not rely on a phase space interpretation. First, we forgo the construction of Floquet bundles and implement what we call zeroth order asymptotic boundary condition. Second, we construct appropriate phase conditions that eliminate spatial translations and neutral modes at $x=\pm\infty$. We then  solve the resulting boundary-value problem directly using finite differences in $x$ and a pseudospectral method in $y$.

\paragraph{Outline.}
The remainder of this paper is organized as follows. In Section \ref{s:1}, we characterize grain boundaries and motivate our numerical approach, which combines a far-field-core decomposition with a domain truncation using zeroth order asymptotic boundary conditions. We refer to an appendix for a more detailed  justification. Section \ref{s:4} is concerned with the numerical implementation of this truncated problem. In particular, we give numerical evidence for convergence as predicted in typical cases. In Section \ref{s:5}, we use the algorithm to study various phenomena associated with grain boundaries. We conclude with a discussion of other potential applications and extensions. 

\begin{Acknowledgment}
D.L. acknowledges funding through the Institute for Mathematics and its Applications and the Faculty Research Support Fund (University of Surrey). A.S. acknowledges partial support from NSF under grants DMS-0806614 and DMS-1311740, support through a DAAD Faculty Research Visit Grant and a WWU Fellowship. A.S. and D.L. acknowledge support from the London Mathematical Society through a Research in Pairs Grant 41502. 

The authors would like to thank J. Lega for many helpful conversations and comments on early versions of this manuscript, and for sharing her preliminary results with us.  The authors also acknowledge discussions with D. Avitabile on numerical aspects of our approach. 
\end{Acknowledgment}

\section{Continuing grain boundaries }
\label{s:1}
In this section, we characterize grain boundary as heteroclinic orbits, not necessarily close to onset. We then describe the inherent difficulties involved with computing grain boundaries in large boxes before laying out our approach via a far-field-core decomposition.

\paragraph{Characterizing grain boundaries.}

We start with a conceptual definition of grain boundaries. We fix $\mu$, throughout, and consider only orientation of grains as free parameters. We first assume the existence of a family of striped solutions $u_\mathrm{s}(kx;k)$, $k\in (k_\mathrm{min},k_\mathrm{max})\supset (k_\mathrm{zz},k_\mathrm{eck})$. A \emph{grain boundary} is a solution $u_*(x,y)$ which converges towards stripes of different orientations as $x\to\pm\infty$. We focus throughout on resonant angles, where the stripes at $\pm\infty$ possess wave vectors $\underline{k}^{\pm}=(k_x^\pm,k_y^\pm)$ that satisfy  $k_y=k_y^+/q_+=k_y^-/q_-$ for integer $q_\pm$. We moreover assume minimal period in the $y$-direction (although this assumption can easily be removed). In summary, we require that $u_*(x,y)$ solves the stationary Swift-Hohenberg equation \eqref{e:sh}, with
\begin{itemize}
\item \emph{periodicity:} $u_*(x,y)=u_*(x,y+L_y)$, $L_y=2\pi/k_y$;
\item \emph{convergence:} $|u_*(x,y)-u_\mathrm{s}(k_x^\pm x+k_y^\pm y;k^\pm)|\to 0$, for $x\to\pm\infty$, uniformly in $y$;
\end{itemize}
here,  $(k^\pm)^2=(k_x^\pm)^2+(k_y^\pm)^2$. Convergence and periodicity imply the resonance condition  $k_y=k_y^+/q_+=k_y^-/q_-$. Uniform convergence is equivalent to convergence of derivatives, using the regularizing properties of the equation. In the following, we will use the rescaled variable $k_y y = :\tilde{y}$, in which we have $2\pi$-periodicity and convergence to $u_\mathrm{s}(k_x^\pm x+q_\pm \tilde{y};k^\pm)$, respectively. The rescaled equation for grain boundaries is, dropping the tildes for the $y$-variable, 
\begin{equation}\label{e:gbr}
-(\partial_x^2+k_y^2\partial_y^2+1)^2u + \mu u - u^3=0.
\end{equation}
With the results in the small-amplitude limit, one expects such solutions to be locally unique up to translations in $x,y$. In particular, for any fixed $k_y$, there exist \emph{selected} wavenumbers $k_x^\pm$ at $\pm\infty$, for which a grain boundary exists. Equivalently, one finds a relation between the angles $\phi_\pm$, depicted in Figure \ref{f:gbsc}, and the selected wavenumbers ${k}^\pm$. 

\paragraph{Computing grain boundaries --- the large box and its problems.}

A first naive approach to solving \eqref{e:gbr} would be to impose periodic boundary conditions on $(x,y)\in (-L_x,L_x)\times (0,2\pi)$. With periodic boundary conditions, one would effectively compute a pair of grain boundaries, possibly located at $x=0$ and $x=L_x$, respectively. One can then try to compute grain boundaries from an initial guess using Newton's method. A first difficulty is caused by the translations in $x$ and $y$, which yield non-uniqueness and a two-dimensional kernel of the linearization at such a solution. One would usually add appropriate phase conditions to eliminate these translations and  add  drift speeds in the $x$- and $y$-direction to set up a well-posed problem, expecting that drift speeds vanish at solutions,
\begin{align}
-(\partial_x^2+k_y^2\partial_y^2+1)^2u + c_x \partial_x u + c_y  \partial_y u +  \mu u - u^3&=0,\quad (x,y)\in (-L_x,L_x)\times (0,2\pi) +\mbox{``periodic'' b.c.},\label{e:box1}\\
\int_{x,y} (u-u_\mathrm{old}) \cdot \partial_x u_\mathrm{old} &=0,\label{e:box2}\\
\int_{x,y} (u-u_\mathrm{old}) \cdot \partial_y u_\mathrm{old} &=0.\label{e:box3}
\end{align}
It turns out that this somewhat standard approach to computations of patterns in periodic domains is viable here only for moderate sizes of $L_x$. Since the solution consists roughly of striped patterns in most of the domain, the linearization of \eqref{e:box1} at a solution will resemble the linearization at a striped pattern throughout most of the domain. For spectrum near the origin, this linearization is well approximated by the Laplacian from the phase-diffusion approximation (or a similar elliptic operator from the Cross-Newell equation). It will therefore inherit spectrum $\lambda_j\sim- j/L_x^2$, $j\in\N$, which accumulates at the origin for $L_x$ large; see \cite{ssabs,radss} for a general treatment of the behavior of continuous spectra under truncation of the domain. The phase conditions \eqref{e:box2}--\eqref{e:box3} can eliminate two neutral eigenvalues but do not resolve the ill-posedness as $L_x\to\infty$. 

\begin{figure}[h]
	\centering
	\includegraphics[width=0.55\linewidth]{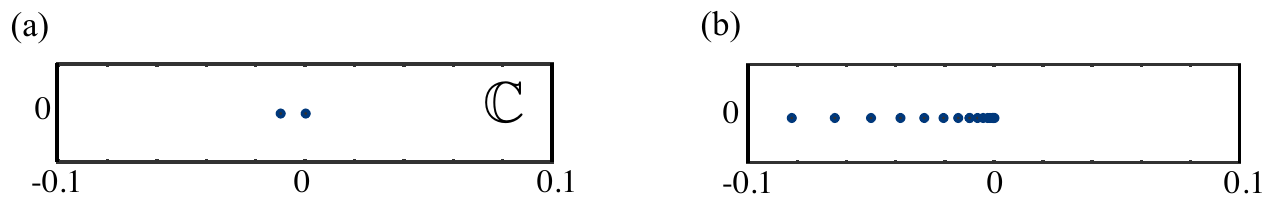}
	\caption{Plot of the eigenvalues in $[-0.1,0.1]$ of the doubly-Fourier discretization of the Swift-Hohenberg equation about $u_s(y;1)$ with (a) ($N_x=20,L_x=10$) and (b) ($N_x=200,L_x=100$) and $N_y=20$. We see that for the same stepsize in $x$, as $L_x$ is increased there is an accumulation of eigenvalues at zero.}\label{f:abs}
\end{figure}

As a consequence, performance of Newton iterations deteriorates with increasing $L_x$. From this perspective, it is clear that this difficulty cannot be eliminated by the choice of separated boundary conditions, such as, say, oblique boundary conditions at $x=\pm L_x$, as used in \cite{ercolani2003}. Figure \ref{f:abs} illustrates this accumulation of eigenvalues near the origin and the resulting ill-posedness.

\paragraph{A remedy: far-field-core decomposition and asymptotic boundary conditions.}

A remedy to the presence of a family of neutral modes is an a priori ansatz for the solution in the far field. We explain the main strategy here and refer to the appendix for more details. One can verify that grain boundaries converge exponentially towards striped patterns, suggesting  a decomposition of the solution via
\[
u(x,y)=w(x,y)+\chi_+(x)u_+(x,y)+\chi_-(x)u_-(x,y),\quad u_\pm(x,y)=
u_\mathrm{s}(k^\pm_x x+q_\pm y+\varphi^\pm;k^\pm),
\]
with smooth cut-off functions 
\[
\chi_\pm(x)=1,\ \pm x>d+1,\qquad 
\chi_\pm(x)=0,\ \pm x<d;
\]
see also \cite{morrissey2015} for a similar approach.
Substituting this ansatz into the Swift-Hohenberg equation, we find
\begin{equation}\label{e:fc0}
\mathcal{L}(w+\sum_\pm \chi_\pm u_\pm)-(w+\sum_\pm \chi_\pm u_\pm)^3=0,\quad \mathcal{L}=-(\partial_x^2+k_y^2\partial_y^2+1)^2+\mu,
\end{equation}
which can be written, after subtracting the equation for $u_\pm$, in the form
\begin{equation}\label{e:fc1}
\mathcal{L}w- \left\{\left(w+\sum_\pm \chi_\pm u_\pm\right)^3-\left(\sum_\pm \chi_\pm u_\pm\right)^3\right\}+\sum_\pm\left[\mathcal{L},\chi_\pm\right]u_\pm 
+\left\{\sum_\pm\chi_\pm u_\pm^3-\left(\sum_\pm\chi_\pm u_\pm\right)^3  \right\}=0,
\end{equation}
where we used the commutator notation $[A,B]u=A(Bu)-B(Au)$. The expression in the last bracket can be viewed as a commutator between nonlinearity and cut-off functions, evaluated on stripe solutions. Note that the residual of \eqref{e:fc1} is exponentially localized when $w$ is, since commutators vanish for $|x|$ large. One may therefore expect that boundary conditions at finite $x=\pm L_x$ only contribute exponentially small corrections $\rmO(\rme^{-\eta |L_x|})$ to the profile $w$ and wavenumbers $k_x^\pm$ and $k_y$.

Given that we are looking for  $w$ to be exponentially localized, Dirichlet boundary conditions  $w=w_{xx}=0$ at $x=\pm L_x$ appear to be a natural choice. Since neither $k^\pm$ nor $\varphi^\pm$ are known, they appear as additional free variables in the equation. Inspecting the geometry of a grain boundary, one readily sees that one can fix $\varphi^\pm=0$, after appropriate shifts in $x$ and $y$. From the point of view taken above, the equations  $\varphi^\pm=0$ act as a phase condition normalizing $x$- and $y$-translations. 

The remaining additional variables $k^\pm$ need to be compensated for by additional equations that eliminate multiplicities. Indeed, fixing $\varphi^\pm$ only eliminates translations of the solution if exponential localization of $w$ is enforced: otherwise, the difference between a grain boundary and its translates can simply be added to $w$. In a bounded domain, however, exponential localization cannot be strictly enforced since weighted and unweighted norms are equivalent. One therefore needs to add a condition on $w$ that eliminates asymptotics $w\sim \partial_x u_\pm$ for $x\sim L_x$. Our choice is 
\begin{equation}\label{e:ph12}
\int_{y=0}^{2\pi}\int_{\pm x=L_x-2\pi/k_x^\pm}^{L_x} w(x,y)\cdot \partial_{\xi^\pm} u_\mathrm{s}(\xi^\pm;k^\pm)\rmd x\,\rmd y =0,
\end{equation}
where $\xi^\pm=k^\pm_x x +q_\pm y$. 
As common with phase condition, the precise form of the condition is not crucial, but averaging over roughly a period appeared to work well. 

From a different point of view, enforcing exponential localization of $w$ is a \emph{zeroth order asymptotic boundary condition}. In computations of homoclinic and heteroclinic orbits, one usually tries to use \emph{first order asymptotic boundary conditions}, approximating the stable manifold by its tangent space. However, boundary conditions in the form of an affine subspace \emph{transverse} to the unstable subspace at the asymptotic profile also give convergence as $L_x\to\infty$, with half the exponential rate \cite{beyn1990}, thus necessitating roughly twice the domain size $L_x$. We refer to such transverse subspaces as zeroth order asymptotic boundary conditions. 

In our case, the periodic orbits come in a two-parameter family, parameterized by $k_x$ and $\varphi$, so that one would wish to approximate a strong stable subspace of the linearization. The computation of the strong stable subspace could prove quite cumbersome. One would need to construct the strong stable and center-unstable adjoint Floquet bundles to  $\mathcal{L}-3u_\pm^2$, written as a first-order evolution operator in $x$ as in \eqref{e:ds}, and the associated spectral projection, which would typically be nonlocal in $y$. While such asymptotic Floquet boundary conditions have been successfully implemented in an ODE contexts \cite{beyn1990,pampel2001,dieci2004,krauskopf2007,doedel2008a,doedel2008b}, we believe that the computational overhead would not outweigh the gain of a factor two in domain size in our case.  

Our choice of Dirichlet boundary conditions  together with the phase condition \eqref{e:ph12} can be seen as a naive construction of a subspace transverse to the center-unstable subspace, which turns out to perform well in most cases. A dimension counting argument, detailed in the appendix, shows that the Dirichlet subspace together with the phase condition yields the correct dimension in a Fredholm sense, so that one may expect transversality to be generic and to fail only at a discrete set of angles (which of course still is a serious concern).

Summarizing, we solve 
\begin{align}
\mathcal{L}\left(w+\sum_\pm \chi_\pm u_\pm\right)-\left(w+\sum_\pm \chi_\pm u_\pm\right)^3&=0,\qquad 
(x,y)\in (-L_x,L_x)\times (0,2\pi) \label{e:bvp1}\\
w=w_{xx}&=0, \qquad (x,y)\in \{-L_x,L_x\}\times (0,2\pi)\label{e:bvp2}\\
\partial_y^j w(x,0) -\partial_y^j w(x,2\pi)&=0,\qquad x\in (-L_x,L_x), \  j=0,\ldots,3,\label{e:bvp3}\\
\int_{x=\pm L_x}^{\pm(L_x-2\pi/k_x^\pm)}\int_{y=0}^{2\pi} u'_\pm w\,\rmd y\,\rmd x&=0, \label{e:bvp4}\\
-\left((k^\pm)^2\frac{\rmd^2}{\rmd \xi^2}+1\right)^2 u_\pm+\mu u_\pm-u_\pm^3&=0, 
\qquad \xi\in (0,2\pi) 
\label{e:bvp5}\\
\frac{\rmd^j}{\rmd \xi^j}u_\pm(0)-\frac{\rmd^j}{\rmd \xi^j}u_\pm(2\pi)&=0, \qquad j=0,\ldots,3, \label{e:bvp6}
\end{align}
where the first equation \eqref{e:bvp1} can also be written in the form \eqref{e:fc1}. We think of this system as an equation in $k_x^\pm$ and $w$, where \eqref{e:bvp5}--\eqref{e:bvp6} are used  for given $k_x^\pm$ (which gives $k^\pm=\sqrt{(k_x^\pm)^2+k_y^2}$) to obtain $u^\pm$, which is then inserted into \eqref{e:bvp1}. 
In the next section, we detail how we discretize this system of equations. In the appendix, we motivate why this decomposition actually gives a well-posed, truncated boundary-value problem, uniformly in $L_x$. 

We also consider the spectrum of the linearization of \eqref{e:bvp1} with respect to $w$, at a solution $w_*, u_\pm^*$
\begin{equation}\label{e:bvplin}
\mathcal{L}_* w=\mathcal{L}w-3\left(w_*+\sum_\pm \chi_\pm u^*_\pm\right)^2 w,\qquad 
(x,y)\in (-L_x,L_x)\times (0,2\pi),
\end{equation}
supplemented with Dirichlet and periodic  boundary conditions \eqref{e:bvp2}--\eqref{e:bvp3} as a rough indicator for temporal stability. We did not attempt to construct asymptotic boundary conditions for the linearization but notice that results as in \cite{ssabs} guarantee that Dirichlet boundary conditions are zeroth order asymptotic outside of the continuous spectrum of the stripes, except for a possibly finite set of eigenvalues $\lambda$.

\section{Implementation and convergence}
\label{s:4}
We describe details of discretization and implementation of the continuation procedure, and demonstrate convergence and robustness of the algorithm.

\paragraph{Discretization and implementation.}

We detail the numerical implementation of the grain boundary problem described in \eqref{e:bvp1}--\eqref{e:bvp6}. The one-dimensional periodic orbits $u_{\pm}$ \eqref{e:bvp5}--\eqref{e:bvp6} are computed on a domain, $\xi\in[0,2\pi)$ with a Fourier pseudo-spectral method; see~\cite{trefethen2000}. In order to interpolate the periodic orbits $u_{\pm}(\xi)$ to the skew coordinates $k_x^{\pm}x+q_\pm y$, we use a band-limited interpolant~\cite[Chapter 3]{trefethen2000}. 

\begin{figure}[h]
	\centering
	\includegraphics{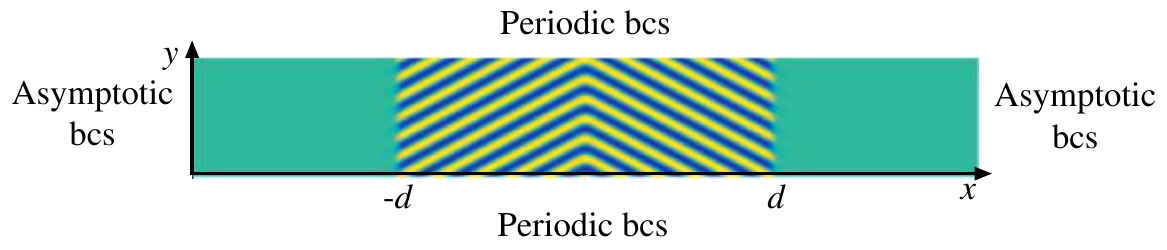}
	\caption{Computational domain for the remainder function $w(x,y)$. We use a Fourier pseudo-spectral discretization in the $y$-direction and a fourth-order finite-differences method in the $x$-direction.\label{f:comp} }
\end{figure}
The computational domain for the remainder function $w(x,y)$ is shown in Figure~\ref{f:comp}. We use the same Fourier pseudo-spectral discretization in the $y$-direction and a standard fourth-order finite-differences methods (see~\cite{leveque2007}). We take the cut-off functions $\chi_{\pm}$ to be $\chi_+(x) = (1+\mbox{tanh}(m(x-d)))/2$ and $\chi_-=1-\chi_+$. The integral phase conditions \eqref{e:bvp4}, are computed using a trapezoidal rule in both $x$ and $y$. The Jacobian for the (now algebraic) system of equations, is explicitly computed with respect to the remainder function $w(x,y)$ and a first order finite-difference is used for the Jacobian with respect to the asymptotic wave numbers $k_x^{\pm}$. The nonlinear algebraic system is then solved for $(w,k^+_x,k^-_x)$ using a trust-region Newton method~\cite{coleman1996}. Parameter exploration is carried out using secant pseudo-arclength continuation~\cite{krauskopf2007}. 

The scheme is implemented in \textsc{matlab} (version 2014b) where typical discretizations in $x$ are $N_x=1000$ mesh points on $x\in[-40\pi,40\pi]$ and $N_y=40$ Fourier collocation points in the $y$-direction. For the cutoff functions, typical values are $m=1,d=100$.

As starting conditions for Newton iterations we used sharp interface grain boundaries, that is, stripes of piecewise constant orientation.

Temporal stability of the grain boundaries is calculated by computing eigenvalues of the linear operator defined  in \eqref{e:bvplin}. We use the same spatial discretization as for the computation of the grain boundaries, yielding a large matrix eigenvalue problem that we solve using \textsc{matlab}'s \verb1eigs1 command that uses an Implicitly Re-Started Arnoldi Iteration~\cite{lehoucq1996,sorensen1992}.  

We next show results that illustrate the robustness and convergence of the algorithm. As a measure for convergence, we used the selected wavenumbers $k^\pm$. We noticed those wavenumbers converge as might be expected to the wavenumber at the zigzag boundary; see \S\ref{s:zz}. We therefore computed the  zigzag (transverse) instability of 1D stripes in AUTO07p~\cite{auto07p}. To do this, we solve for the 1D stripes, $u(x)$, and compute the transverse instability criterion, i.e., $\lambda_t$, 
\begin{equation}\label{e:zigzag}
\lambda_t = 2\frac{\langle(\partial_x^2-1)u_x,u_x\rangle}{\langle u_x,u_x\rangle},
\end{equation}
where $\lambda_t$ is the Eigenvalue associated with transverse perturbations of the form $\hat u e^{iky}$. If $\lambda_t<0$, the 1D stripes are transversely stable and the zigzag instability boundary occurs when $\lambda_t=0$; see~\cite{mielke1997}. Setting $\lambda_t=0$ allows one to fix the 1D stripe wavenumber, $k_{\mathrm{zz}}$. We compute the 1D stripes in AUTO to a relative tolerance of $10^{-6}$ and the zigzag criterion boundary to a relative tolerance of $10^{-10}$.

\paragraph{Convergence of the algorithm.}

We present  results on the convergence of the algorithm and its sensitivity to the computational parameters $N_x,N_y,L_x,m$ and $d$. We will also illustrate the effectiveness and limitations of the phase conditions. 

\begin{figure}[h]
	\centering
	\includegraphics[width=0.7\linewidth]{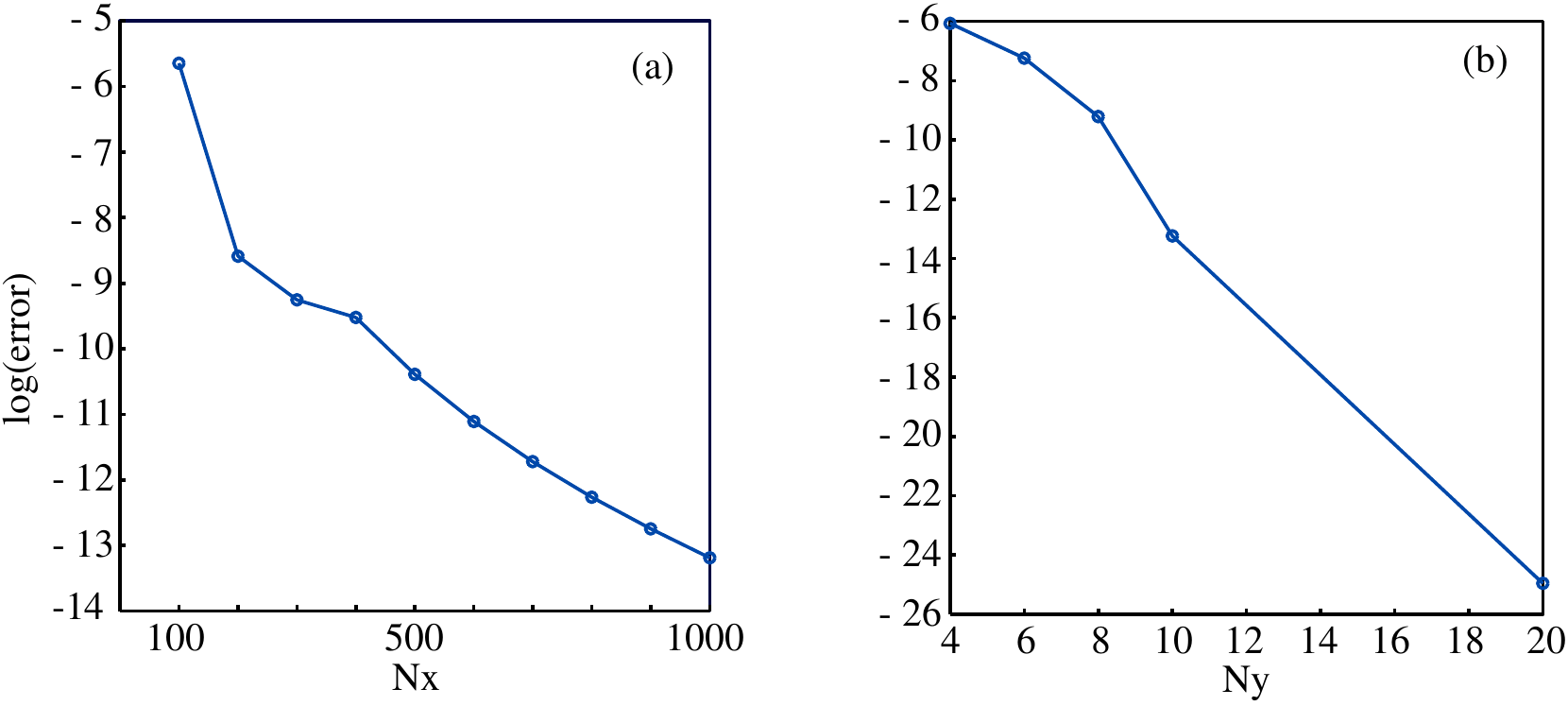}
	\caption{Convergence of the selected wavenumber of the far-field stripes for (a) $\mu=1,k_y=0.85,n_y=40,L_x=80\pi,d=100$, $n_x$ is the number of finite-difference points used in the $x$-direction (b) $\mu=1,k_y=0.85,n_x=1000,L_x=80\pi,d=100$. Error defined as the difference from the selected wave number with $\mu=1,k_y=0.85,n_y=40,n_x=1000, L_x=80\pi,d=100$.\label{f:convergence1}}
\end{figure}

Our test case will be a weakly bent symmetric grain boundary at $\mu=1$ in~(\ref{e:sh}) where we take $k_y=0.85$. The selected asymptotic wavenumber $k^{\pm}$ of the grain boundary is the critical zigzag instability wavenumber of the 1D stripes i.e. $k_\mathrm{zz}=0.9991$. In Figure~\ref{f:convergence1}, we plot the difference (error) between the computed asymptotic wave number and the zigzag wavenumber $k_\mathrm{zz}$. 
We see in Figure~\ref{f:convergence1}, that even for rather crude discretizations the asymptotic wave number of the stripes of the grain boundary is very well approximated. We see in particular spectral (geometric) convergence as we increase the number of Fourier collocation points $N_y$.

\begin{figure}[h]
	\centering
	\includegraphics[width=0.7\linewidth]{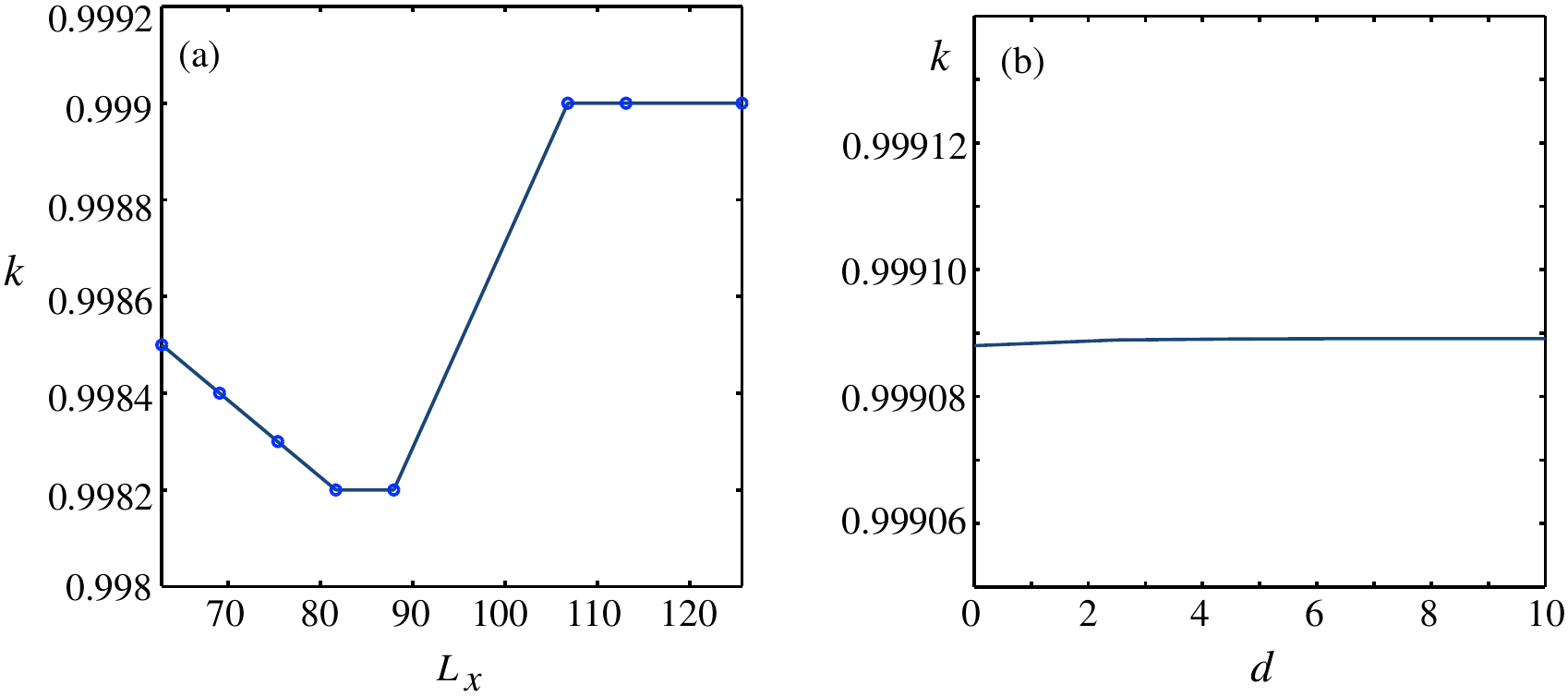}
	\caption{Convergence of the selected wavenumber of the far-field stripes for (a) $\mu=1,k_y=0.85,n_y=40,d=50, n_x=500$, varying $L_x$ and (b) varying the cutoff $d$ with $L_x=40\pi$.\label{f:convergence2}}
\end{figure}
In Figure~\ref{f:convergence2}, we show how the selected asymptotic wavenumber $k^{\pm}$ depends on $L_x$ and the cutoff distance $d$. We see in Figure~\ref{f:convergence2}(a) that, if $L_x$ is greater than about twice the cutoff distance $d$, then the far-field selected wavenumber is independent of $L_x$ (here $k_y=0.85$). In particular, we find that the length of the domain needs to be sufficiently large such that the remainder function ,$w$, is zero for one far-field skewed stripe. However, we see that for even small $L_x$ the selected far-field wavenumber is reasonably accurate. In Figure~\ref{f:convergence2}(b) we see that the cutoff distance, $d$ has almost no affect on the selected wavenumber, $k$;  see for instance~\cite{beyn1990} for results of exponential convergence in $L_x$ of heteroclinic orbits. 

Next we compute the condition number of the system \eqref{e:bvp1}--\eqref{e:bvp6} about $w(x,y)$ as we vary the discretization in $x$. We use \textsc{matlab}'s \verb1condest1 routine~\cite{hager1984} to compute a lower bound for the 1-norm condition number. We find that the condition number estimate is of order $10^9$ for typical discretizations (significantly less than machine precision) and robust with respect to changes in the computational parameters. 

\begin{figure}[h]
\centering
\includegraphics{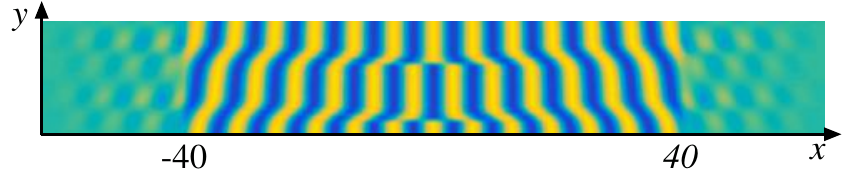}
\caption{Plot of $w(x,y)$, in the situation where we  observe the emergence of non-vanishing tails outside $x\in[-d,d]$; $\mu=1,k_y=9.636267\times10^{-2},n_y=20,n_x=500,L_x=40\pi,d=40$.\label{f:loss_transverse}}
\end{figure}
The rate of exponential convergence is related to the temporal stability of the stripes through the complex dispersion relation, as we will explain in the appendix. Indeed, for both horizontal rolls and for large vertical periods, $k_y\ll 1$, the zigzag instability manifests itself via slowly decaying tails of $w$. Figure ~\ref{f:loss_transverse}, shows the remainder function, $w(x,y)$ for $k_y=9.636267\times10^{-2}$. Note in particular the tails in $w(x,y)$ outside the cut-off window $x\in[-d,d]$.

\section{Applications}
\label{s:5}

We apply the numerical procedures outlined above to study grain boundaries in the Swift-Hohenberg equation \eqref{e:sh}. Fixing the parameter $\mu$, we continue grain boundaries in the angle and study their properties and possible bifurcations. We start by investigating wavenumber selection, in particular the fact that grain boundaries tend to select marginally stable stripe patterns at the zigzag boundary in Section \ref{s:zz}. After briefly discussing phase selection at the interface, Section \ref{s:phase}, we focus on the behavior of grain boundaries as the angle is varied from obtuse to acute, Section \ref{s:path}. We then study grain boundaries with $(q_-,q_+)$ different from $(1,-1)$, exhibiting an interesting bifurcation near grain boundaries interfacing horizontal stripes in Section \ref{s:hor}. Finally, we show pinning effects in Section \ref{s:other}, when the core of grain boundaries widens to contain patches of vertical stripes. We also show that grain boundaries between spots are significantly more complicated, with more dominant pinning effects leading to snaking bifurcation diagrams. 

\subsection{Selection of marginally stable stripes}\label{s:zz}
\begin{figure}[h]
	\centering
	\includegraphics[width=0.6\linewidth]{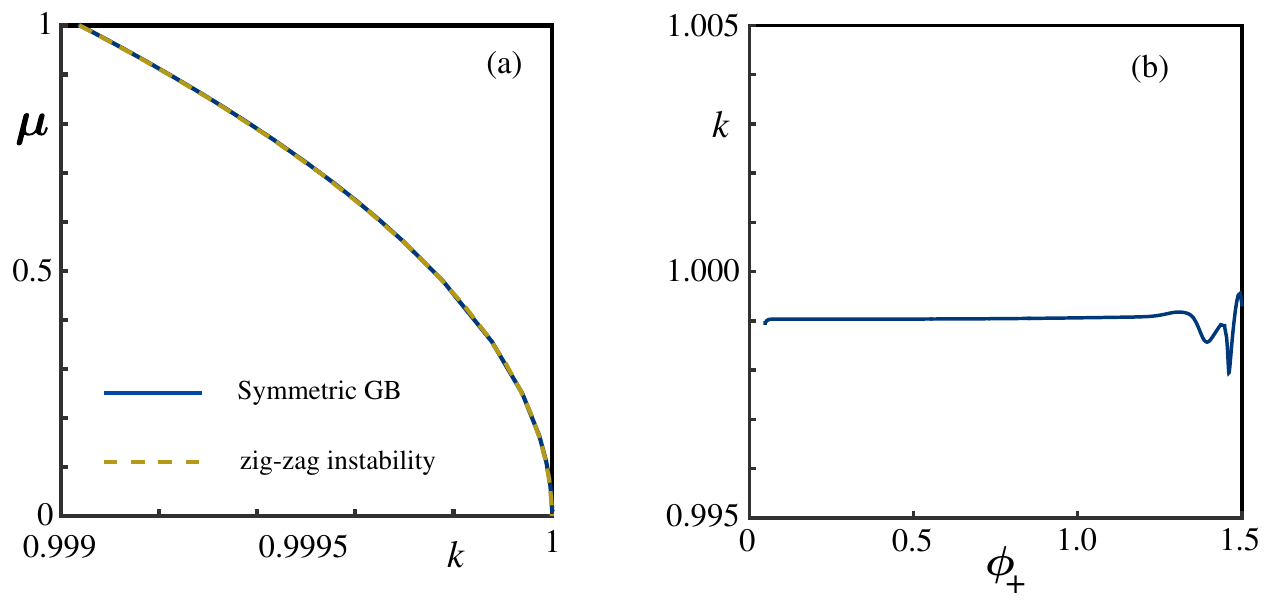}
	\caption{Wavenumber $k$ of stripes selected by the symmetric grain boundary vs $\mu$ is varied, angle fixed at $\phi_+=0.5534$, $k_y=0.85$, (a) and wavenumber $k$ vs.  angle $\phi_+=\mbox{arccos}(k_y/k)$ (compare Fig. \ref{f:gbsc}), parameter $\mu=1$ fixed. Computational parameters are $n_x = 600,L_x=10\pi,n_y=20$.}
	\label{f:symmetric_gb_k}
\end{figure}

At small amplitude, $\mu\ll 1$, the leading-order description of grain boundaries via the amplitude equation shows that grain boundaries select wavenumbers, that is, for a fixed angle, grain boundaries exist only for a particular wavenumber in the far field. at leading order in $\mu$, this wavenumber is $k=1$ in the Swift-Hohenberg equation. We demonstrate here numerically that this property holds also at finite amplitude, and show that the selected wavenumber agrees with the wavenumber defined by the zigzag boundary in~(\ref{e:zigzag}); see Figure \ref{f:symmetric_gb_k}. The numerical discrepancy is less than $10^{-10}$. For very acute angles, that is, large vertical period $2\pi/k_y$, the resolution in $y$ is poor and we observe discrepancies. For very obtuse angles, weak bending, convergence of grain boundaries to stripes is slow in $x$ \cite{haragus2007} which also introduces numerical inaccuracies.

\paragraph{Variational reasons for marginal stability.}
A possible reason for the selection of marginally zigzag stable stripes can be seen by looking at the energy of the Swift-Hohenberg equation. The Swift-Hohenberg equation~(\ref{e:sh}) is a gradient flow 
\[
u_t = -\nabla\mathcal{E}(u),
\]
in $H^2(\R^2)$, where the energy functional $\mathcal{E}$ is given by
\begin{equation}\label{e:energy_SH}
\mathcal{E}(u) = \int_{\R^2}\left[\frac{1}{2}[(1+\Delta)u]^2 -\frac{1}{2} \mu u^2+ \frac{1}{4}u^4\right]\mbox{d}\mathbf{x},\qquad \mathbf{x}\in\R^2,
\end{equation} 
and the gradient $\nabla\mathcal{E}(u) = \frac{\delta\mathcal{E}}{\delta u}(u)$ of $\mathcal{E}$ with respect to $u$ is computed with respect to the $L^2(\R^2)$ inner product. Equilibria of the Swift-Hohenberg equation are critical points of the energy functional. Since grain boundaries appear to be stable, it is natural to look for grain boundaries as local minimizers of $\mathcal{E}$. To our knowledge, an existence proof that constructs grain boundaries as minimizers using methods from the calculus of variations is not available, even looking beyond the example of the Swift-Hohenberg equation. 

On the other hand, one can envision initializing a system with an interface between stripes of different orientation as we did in Figures \ref{f:sn} and \ref{f:cont}, which leads to mixing of wavenumbers in the far field. Since in this process, the energy of the system  decreases, one concludes that the wavenumber selected by the grain boundary necessarily should correspond to the wavenumber of stripes with minimal energy per unit length. More precisely, one can define the average energy of a stripe $u_\mathrm{s}(\xi;k)$, $\xi=kx$, as
\[
\mathcal{E}(k) = \frac{1}{2\pi}\int_0^{2\pi}\left[\frac12[(1+k^2\partial_\xi^2)u]^2 - \frac12\mu u^2 + \frac14u^4\right] \rmd \xi,
\]
and minimize with respect to $k$. The minimum is then attained at the zigzag boundary, $k=k_\mathrm{zz}$. Renormalizing the energy, 
\begin{equation}\label{e:energy_SHr}
\mathcal{E}_\mathrm{re}(u) = \int_{\R^2}\left[\frac{1}{2}[(1+\Delta)u]^2 -\frac{1}{2}\mu u^2 + \frac{1}{4}u^4-\mathcal{E}(k_\mathrm{zz}) \right]\mbox{d}\mathbf{x},\qquad \mathbf{x}\in\R^2,
\end{equation} 
one expects to find a local minimum at a grain boundary at finite energy when restricting to functions with periodicity $2\pi/k_y$ in $y$, and evaluating integrals on a fundamental domain.

\paragraph{Hamiltonian reasons for marginal stability.}

It turns out that, given existence of a grain boundary, one can use the variational structure to conclude that grain boundaries select the zigzag critical wavenumber, exploiting the fact that the spatial dynamics formulation \eqref{e:ds} defines an ill-posed Hamiltonian equation. As a consequence of Noether's theorem, the equation then possesses conserved quantities associated with the continuous symmetries of the equation, namely translations in $x$ and $y$. To be more precise, in the notation from \eqref{e:ds}, consider the symplectic form generated by $L^2$-inner product and the skew-symmetric matrix $J$,
\[
J=\begin{pmatrix}
0&0&0&1\\ 
0&0&-1&0\\
0&1&0&0\\
-1&0&0&0
\end{pmatrix},
\qquad J^T=-J=J^{-1},\ J^2=-\mathrm{id}, 
\]
which, writing $q=(u,u_1)^T$, $p=(v,v_1)^T$, is simply the standard symplectic form, and the Hamiltonian
\[
H[\underline{u}]=\int_y h(\underline{u}),\quad h(\underline{u})=-\frac{1}{2} v^2+u_1v_1 + v(u_{yy}+u)+G(u),\  \underline{u}=(u,u_1,v,v_1)^T,\  G(u)=-\frac{\mu}{2}u^2+\frac{1}{4}u^4,
\]
where we have re-scaled $y$ to be of period $2\pi/k_y$.
Then \eqref{e:ds} can be written in the form
\[
\underline{u}_x=J\nabla_{L^2}H[\underline{u}],
\]
and the Hamiltonian $H$ is conserved. In addition, the translation symmetry in $y$ induces an additional conserved quantity $S$ which we will refer to as momentum,
\[
S[\underline{u}]=\int_y s(\underline{u}),\quad s(\underline{u})=u (v_1)_y+v (u_1)_y,\quad J\nabla_{L^2}S[\underline{u}]=\partial_y\underline{u},
\]
and, in particular, for solutions of \eqref{e:ds},
\begin{equation}
\frac{\rmd}{\rmd x} H[\underline{u}(x,\cdot)]=
\frac{\rmd}{\rmd x} S[\underline{u}(x,\cdot)]=0.
\end{equation}
As a consequence, both Hamiltonian and momentum are equal on asymptotic stripes of grain boundaries, 
Slightly abusing notation, define $H(k):=H[\underline{u}_\mathrm{s}^k]$,  $S(k):=S[\underline{u}_\mathrm{s}^k]$, where $u_\mathrm{s}^k$ is the striped pattern with wave vector $k=(k_x,k_y)$, and write $k^\pm$ for the asymptotic wave vectors of a grain boundary. Further writing $u_\mathrm{s}=u_*(k_x x + k_y y;|k|)$, we obtain
\begin{align}
H(k)&=\int_\xi \left(\frac{1}{2}k_y^4-\frac{3}{2}k_x^4-k_x^2k_y^2\right)(u_*'')^2 + (k_x^2-k_y^2)(u_*')^2+\frac{1}{2}u_*^2+G(u_*),\nonumber\\
S(k)&=2k_xk_y\int_\xi |k|^2 (u_*)''-(u_*')^2.
\end{align}
Marginal zigzag stability occurs when $\int_y |k|^2 \left(u_*''\right)^2-\left(u_*'\right)^2=0$, as one readily verifies by minimizing the energy of stripes. Therefore, $S=0$ precisely when $k=k_\mathrm{zz}$. For symmetric grain boundaries, $k_x^-=-k_x^+$, $k_y^-=k_y^+$ such that $S(k^-)=S(k^+)$ implies $|k^-|=|k^+|=k_\mathrm{zz}$.

One also readily verifies that for $|k|=k_\mathrm{zz}$, 
\[
H(k)=\int_\xi -\frac{1}{2}|k|^4 (u_*'')^2 + \frac{1}{2}u^2 G(u_*),
\]
depends only on $|k|$, implying that arbitrary orientations of marginally zigzag stable stripes are compatible.

\paragraph{Non-variational effects --- selection of stable stripes.}

As noted above, the zigzag boundary is usually associated with an orientational instability induced by the fact that stripes can reduce their local wavelength through local shear in the direction of the wave vector. As a consequence, one notices an instability of stripes with wavenumber smaller than the energy-minimizing zigzag wavenumber. More directly, one sees that the linearization of stripes becomes unstable as the wavenumber $k$ is decreased through $k_\mathrm{zz}$. More precisely, the linearization
\[
\mathcal{L}_\mathrm{s}v=-(\Delta+1)^2 v + \mu v - 3 u_\mathrm{s}(kx;k)v,
\]
can be written in Fourier-Bloch space as 
\[
L_\mathrm{s}(\ell,\sigma;k)w=-\left((\partial_x+\rmi\sigma)^2-\ell^2\right)^2 w +\mu w-3 u_\mathrm{s}(kx;k)w,
\]
where $L_\mathrm{s}$ is posed on $2\pi$-periodic functions, with Fourier-Bloch parameter $\sigma,\ell$. The eigenvalue $\lambda=0$ associated with translation $u_\mathrm{s}'$ at $\sigma=\ell=0$, possesses an expansion 
$\lambda(\sigma,\ell)=-d_{||}(k)\sigma^2-d_\perp(k) \ell^2 + \rmO(4)$, where $d_{||}$ changes sign at $k=k_\mathrm{zz}$ where $d_{||}$ and $d_{\perp}$ correspond to perturbations parallel and perpendicular to the wave vector; see~\cite{cross1993,mielke1997}. In this sense, stripes at the zigzag boundary are marginally stable in the family of stripes. 

\begin{figure}[h]
	\centering
	\includegraphics[width=0.9\linewidth]{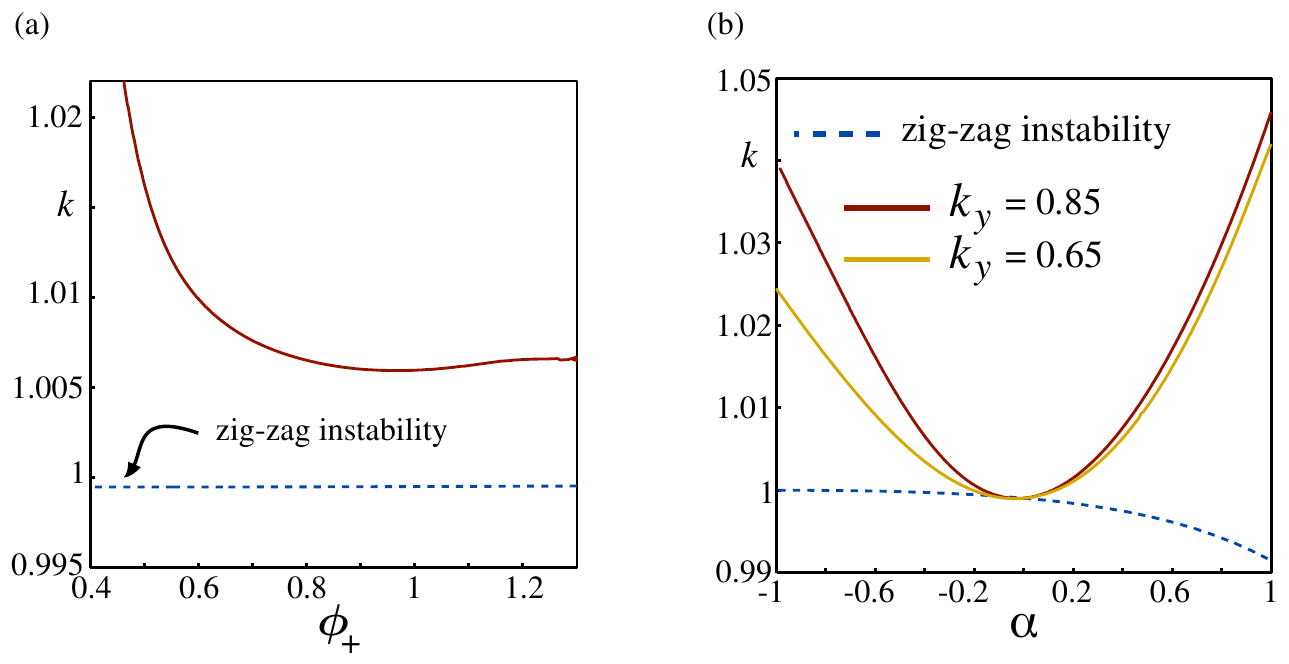}
	\caption{ (a) Selected wavenumber $k$ as a function of the angle of the selected far-field stripes for $\mu=1$ and $\alpha=-0.5$, compared to the wavenumber of zigzag marginally stable stripes.  (b) Selected wavenumber $k$ as a function of $\alpha$ for fixed $k_y=0.85$ and $k_y=0.65$ . Here, a term $+\alpha(\nabla u)^2u$ has been added to the right-hand side of the Swift-Hohenberg equation.\label{f:nonvar_selection}}
\end{figure}

We demonstrate below that the variational characterization of the zigzag boundary is responsible for the selection by grain boundaries, rather than the marginal stability. We therefore perturb the Swift-Hohenberg equation~(\ref{e:sh}), adding $\alpha(\nabla u)^2u$ to the right hand side. As a consequence, variational characterizations of the zigzag boundary are not available anymore. On the other hand, the sign change of $d_{||}$ still occurs at some critical wavenumber $k_\mathrm{zz}(\alpha)$. We computed both this marginally stable  wavenumber $k_\mathrm{zz}(\alpha)$ and the wavenumber selected by the grain boundaries. The results show that grain boundaries always select zigzag stable stripes. Marginally stable stripes are selected only in the variational case $\alpha=0$; see Figure \ref{f:nonvar_selection}. It would be interesting to understand this rigidity theoretically, that is, explain the fact that selected wavenumbers move towards stable stripes upon addition of non-variational terms.

\subsection{Phase selection at grain boundaries --- non-adiabatic effects}\label{s:phase}

\begin{figure}[h]
	\centering
	\includegraphics[width=0.4\linewidth]{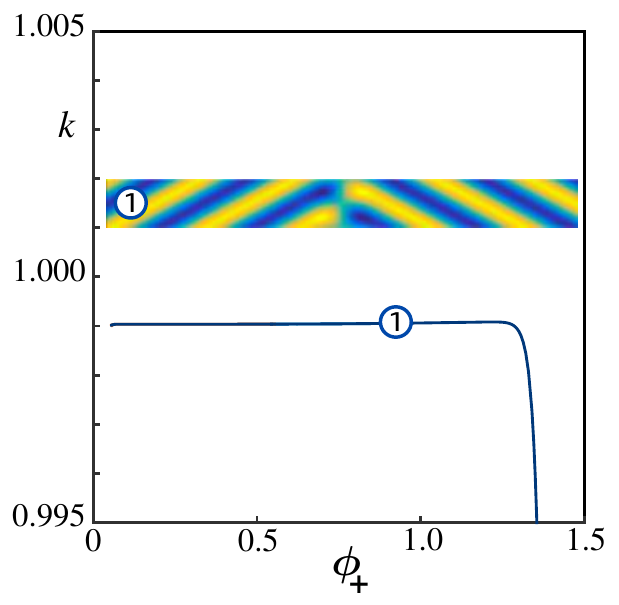}
	\caption{Selected wavenumber of far-field stripes for the odd grain boundary as a function of the angle between stripes with contour plot of computed odd grain boundary as inset.   Computational parameters: $n_x = 1000,L_x=10\pi,n_y=20$; sub-panel is on $(x,y)\in[-30,30]\times[-\pi,\pi)$.}\label{f:anti_phase}
\end{figure}

In the normal form at small amplitudes, there exists a family of grain boundaries, in which stripes at $\pm\infty$ can be shifted vertically relative to each other. One expects this normal form or averaging symmetry to be present at all orders in an expansion, while terms beyond all orders enforce the selection of a relative vertical phase of asymptotic stripes at the grain boundary. In a simplistic picture, one can envision effective gradient dynamics on the circle of grain boundaries parameterized by the relative phase, with at least two critical points. The proofs in \cite{haragus2012} show that even and odd (in $x$) grain boundaries persist, with a phase-mismatch of $0,\pi$, respectively, at $x=0$. We computed odd grain boundaries and showed that they possess properties similar to even grain boundaries, that is, they select zigzag marginally stable stripes; see Figure \ref{f:anti_phase}. Consistent with the simplistic effective dynamics on the circle of relative phases, we find that these anti-phase matched grain boundaries are temporally unstable for all angles. 

\subsection{Acute angles: from grain boundaries to dislocations and disclinations}\label{s:path}

Phenomenologically, one is interested in the behavior of grain boundaries as the angle of asymptotic stripes is changed. This section and the next present a detailed numerical study, continuing grain boundaries in the combined angle. This section is concerned with the simplest case, $(1,-1)$, even grain boundaries. We turn to asymmetric grain boundaries, with different $(q_-,q_+)$, in the next section. 

\begin{figure}[tbhp]
	\centering
	\includegraphics[width=0.75\linewidth]{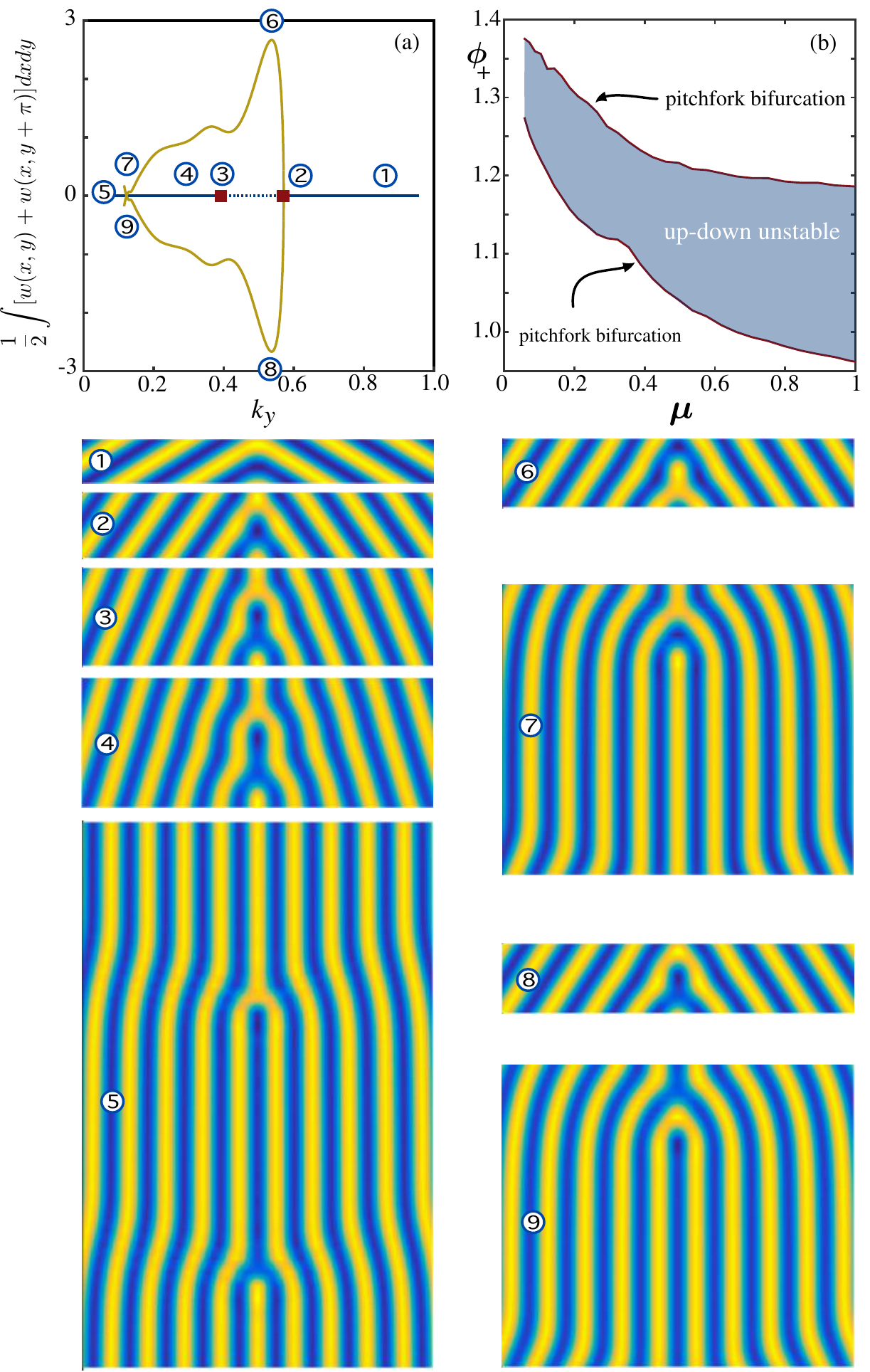}
	\caption{(a) Bifurcation diagram of $(1,-1)$-grain boundaries and (b) instability interval of primary branch as a function of $\mu$ in the shaded region, bounded by $\phi_\mathrm{pf,1}$ (upper red curve) and  $\phi_\mathrm{pf,2}$ (lower red curve). Shown below are several sample profiles along primary and secondary branch.  the primary grain boundary branch showing the first pitchfork branch. Various grain boundary profiles are depicted.\label{f:GB_dis} }
\end{figure}

\paragraph{Continuation to acute angles.}
Recall that symmetric grain boundaries are symmetric with respect to reflections $x\mapsto -x$, and in addition with respect to a parity-shift transformation,  $u\mapsto -u$, $y\mapsto y+\pi$. For obtuse angles $\phi_+$ (compare Figure \ref{f:gbsc} for definition of angles), the results in \cite{haragus2007} show that these are the only possible grain boundaries. We continue this branch for fixed $\mu=1$ in the angle. We do not impose reflection or parity-shift symmetries. Figure \ref{f:GB_dis} shows the results of this computation. We see that the primary branch of symmetric grain boundaries continues to arbitrarily small angles. As we continue, the shape of grain boundaries changes, as a protrusion at the interface develops. Eventually, for acute angles, the symmetric grain boundary consists of a pair of dislocations, conjugate to each other by the parity-shift transformation. We find, however, that this symmetric, primary branch is unstable for a range of angles, against perturbations that break the parity-shift symmetry. The primary branch restabilizes for yet smaller angles. The left panel of Figure \ref{f:GB_dis} shows typical grain boundaries along the primary branch.

\paragraph{Parity-shift breaking bifurcations: destabilization and restabilization of the primary branch.}

The primary branch destabilizes at an angle $\phi_\mathrm{pf,1}(\mu)$ in a parity-shift symmetry breaking pitchfork bifurcation. It remains unstable until a smaller angle $\phi_\mathrm{pf,2}(\mu)$ is reached. At $\phi_\mathrm{pf,1}(\mu)$, grain boundaries with broken parity-shift symmetry bifurcate. We continued this bifurcating branch down to small acute angles and did not detect further bifurcations or instabilities along this asymmetric branch. In particular, the bifurcating branch does not reconnect to the primary branch at $\phi_\mathrm{pf,2}(\mu)$. We suspect that at $\phi_\mathrm{pf,2}(\mu)$ an unstable branch of grain boundaries bifurcates from the primary branch, separating the basins of attraction of the two stable branches for small angles, but we were not able to continue this secondary branch. The right panel in Figure \ref{f:GB_dis} depicts typical profiles along the bifurcated branch. The top panels in Figure \ref{f:GB_dis} show the bifurcation diagram for $\mu=1$, with $k_y\approx0.6$ at the first pitchfork bifurcation ($\phi_\mathrm{pf,1}(\mu=1)\sim 0.962$)  and the instability interval $(\phi_\mathrm{pf,2}(\mu),\phi_\mathrm{pf,1}(\mu))$ for a $\mu\in (0,1)$. We note that, in agreement with the discussion in Section \ref{s:zz}, selected wavenumbers agree with the zigzag stability boundary for both primary and secondary branch.  

Numerics suggest that $\phi_\mathrm{pf,1/2}(\mu)\to \pi/2$ for $\mu\to 0$. This is in agreement with the small amplitude bifurcation analysis in \cite{scheel2014}, where grain boundaries where constructed for $0<\mu<\mu_\mathrm{max}(\phi_+)$, where $\mu_\mathrm{max}(\phi_+)>0$ could converge to zero as $\phi_+\to \pi/2$. In particular, the analysis in \cite{scheel2014} did not suggest any bifurcations along the primary branch of grain boundaries, and we suspect that the bifurcation observed here is outside of the range of validity of the analysis there. 

The parity-breaking bifurcation is related to observations in ~\cite{ercolani2003}.
Ercolani~{\it et al.}~\cite{ercolani2003}, numerically showed how disclinations form at the grain boundaries as the angle between the asymptotic stripes $\phi$ becomes large. We note in particular the result of the direct simulation, reproduced in Figure \ref{f:stadion}, where a family of stripes creates a grain boundary with continuously decreasing angle $\phi_+$. One clearly notices the qualitative change induced by the parity-shift breaking pitchfork bifurcation.

\paragraph{Defects along primary and secondary branches.}

\begin{figure}[t]
	\centering
	\includegraphics[width=0.75\linewidth]{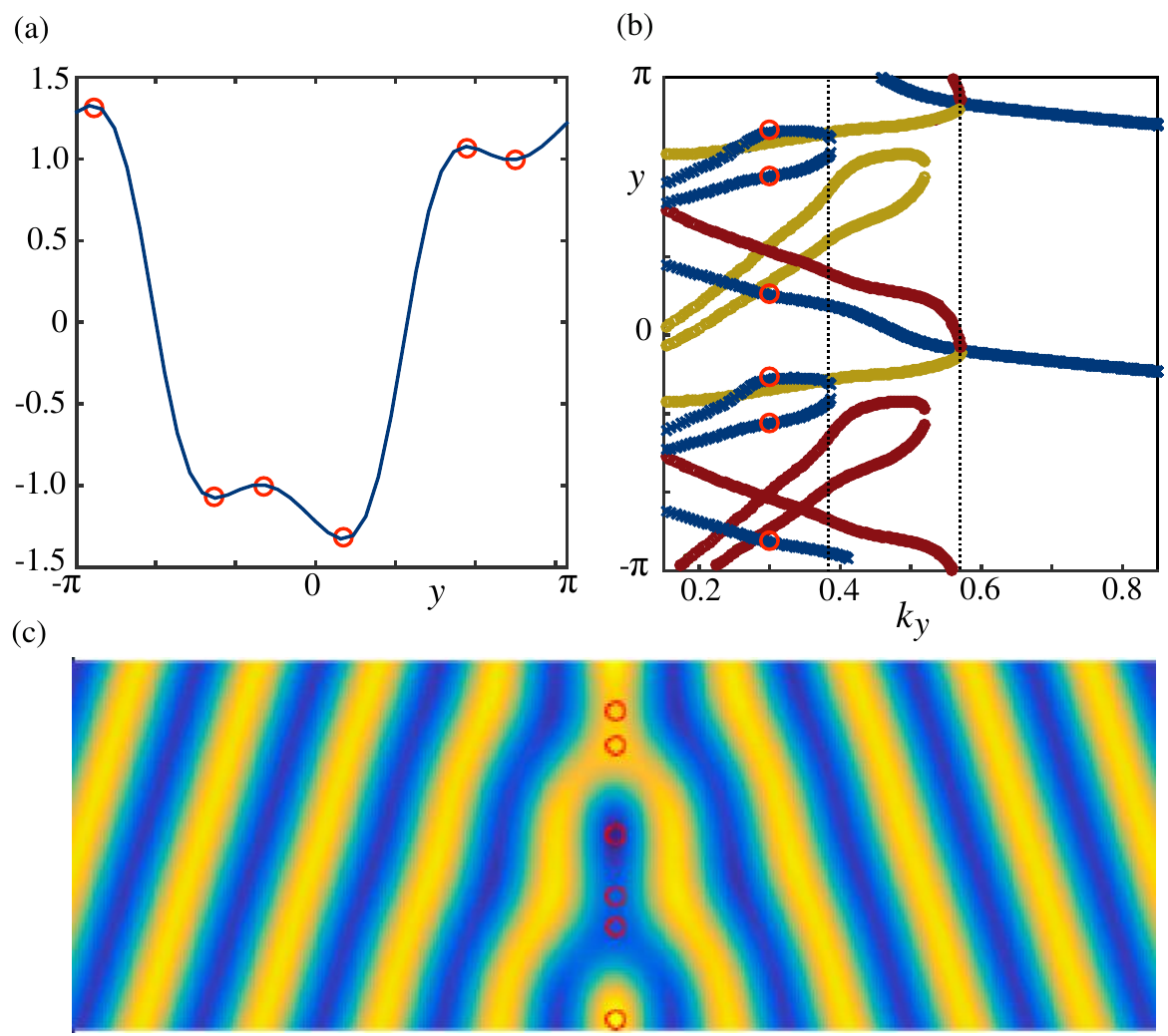}
	\caption{(a) Plot of $u(0,y)$ of the primary grain boundary at $k_y=0.3$, with turning points indicated by circles ;(b) plot of defect locations as $k_y$ is varied for both primary (blue) and secondary branches (gold and red, resp.). The vertical dotted lines denote the locations of the pitchfork bifurcations at $\phi_\mathrm{pf,1}$ and $\phi_\mathrm{pf,2}$; (c) plot of the primary grain boundary at $k_y=0.3$ with defect locations indicated by circles corresponding to (a).\label{f:defect}}
\end{figure}

\begin{figure}[h]
	\centering
	\includegraphics[width=0.7\linewidth]{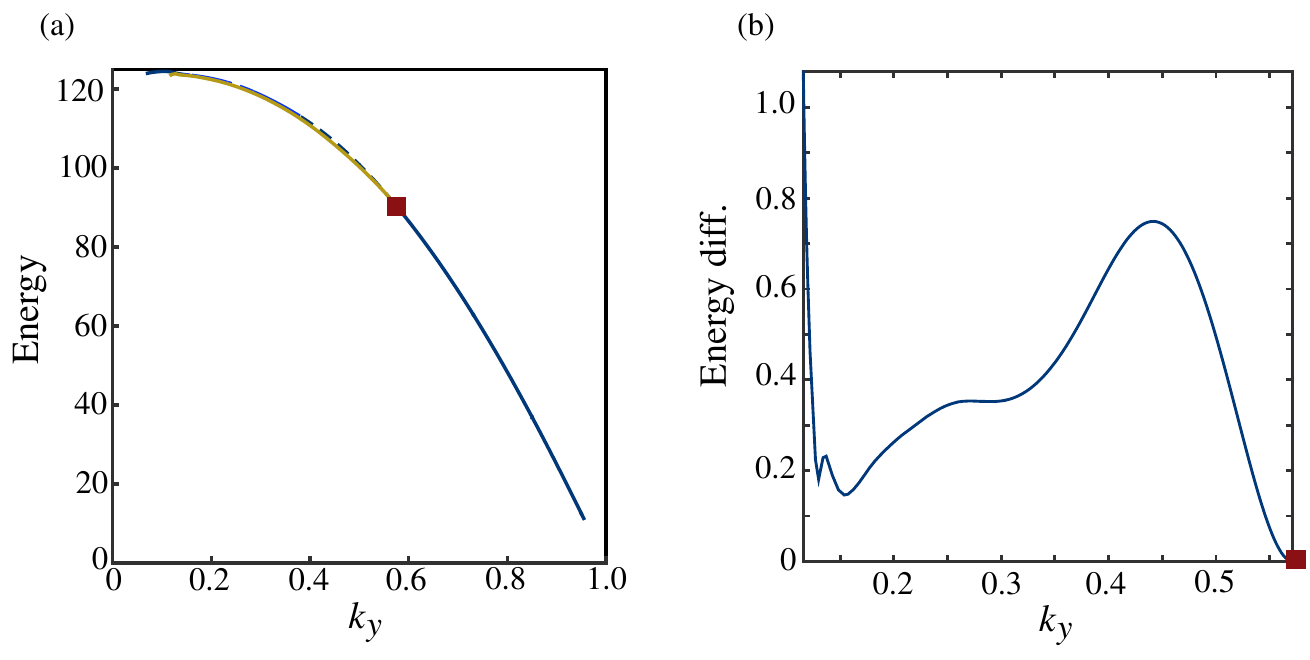}
	\caption{(a) Energy of primary and secondary symmetric branches; (b) energy difference between primary (higher energy) and secondary (lower energy) branches.\label{f:energy1}}
\end{figure}

At least phenomenologically, grain boundaries can be described in terms of defects that may or may not be created when stripes of different orientation create interfaces. The work in \cite{passot1994,ercolani2003,ercolani2009} provides a rationale for the emergence of defects in a largely model independent context, relying rather on the description of stripes via phase modulation equations, and the crucial fact that phase gradients may exhibit apparent singularities since wave vectors are directors rather than vectors due to the underlying reflection symmetries of stripes. 

Phenomenologically, one notices in Figure \ref{f:GB_dis} that, as the angle $\phi_+$ is decreased, distinct point defects develop at the grain boundary. Along the primary branch, a pair of dislocations, conjugate by the parity-shift symmetry, develops. Along the asymmetric branch, these two conjugate dislocations split into disclinations, two of which cancel, the other two forming a bound state; see Figure \ref{f:schematic}, below for more schematic depictions of grain boundaries. 

We therefore try to track the emergence of defects at the grain boundary in this particular example of the Swift-Hohenberg equation. A difficulty one faces in such characterizations is that the location, or even existence of a point defect is not universally defined, rendering the preceding phenomenological discussion imprecise. One usually looks for a singularity of the phase or its gradient, which however requires an unambiguous definition of the phase almost everywhere. 

Here, we define (somewhat arbitrarily) a defect of an even grain boundary to be a critical point of the profile $u(0,y)$; see  Figure~\ref{f:defect}(a) \& (c). Note that such critical points correspond to critical points of $u(x,y)$ since $u(x,y)=u(-x,y)$ such that $x$-derivatives vanish at $x=0$. Figure~\ref{f:defect}(b), shows defects of primary and secondary branches as we vary the angle (alias $k_y$). Note that we also track the global maximum and minimum, which exists also for obtuse angles due to periodicity, but our interest is in newly emerging critical points. 

We observe that the primary branch develops four  defects (two pairs, conjugate by parity-shift symmetry) just before the re-stabilizing pitchfork at $k_y\approx0.4$. For the secondary pitchfork branches, we observe that maximum and minimum continue from those of the primary branch at $k_y\approx0.6$. Later, two new defects develop at $k_y\approx0.5$. We found further critical points for small $k_y$ but chose not to indicate them in Figure~\ref{f:defect}(b) as second derivatives were small at these points, indicating that they do not give well defined defect locations.

\paragraph{Energy of grain boundaries along primary and secondary branches.}

In the region of bistability, one can attempt to derive a selection criterion for grain boundaries based on the energy. Since grain boundaries converge exponentially to stripes with energy-minimizing wavenumber, they possess finite renormalized energy $\mathcal{E}_\mathrm{re}$ as defined in \eqref{e:energy_SHr}. We computed the energy of the  marginally stable stripes $\mathcal{E}_\mathrm{zz}$ using AUTO07p and used this result to compute the renormalized energy of the grain boundaries. The results are shown in Figure~\ref{f:energy1}. As expected, we notice that for weak bending, that is, obtuse angles, $\phi_+\to \pi/2$, $k_y \to k_\mathrm{zz}$, the energy tends to zero. The energy increases monotonically as the angle is decreased. Energies of primary and secondary branch differ very little. In the bistability region for $k_y\lesssim 0.6$ the energy of the secondary branches is slightly lower than the energy of the primary branch thus indicating a weak preference for parity-shift broken grain boundaries at small angles.

\subsection{Asymmetric grain boundaries --- small resonances}\label{s:hor}

Generally, grain boundaries involve two angles $\phi_+$ and $\phi_-$  of stripes relative to the grain boundary interface. In this regard, the grain boundaries considered thus far are a very special subclass. We  describe here  how to study asymmetric grain boundaries, $\phi_+\neq \phi_-$, with several restrictions. First,  our approach is tied to the case of resonant angles. Second, resonances where $q_\pm$ are large require small $k_y$, or, equivalently, large $y$-periods, which increases computational cost significantly. We therefore limit ourselves to resonances where $q_\pm$ are small. The results do suggest however a building pattern for resonant grain boundaries with large $q_\pm$ or even non-resonant grain boundaries.

\begin{figure}[h]
	\centering
	\includegraphics[width=0.7\linewidth]{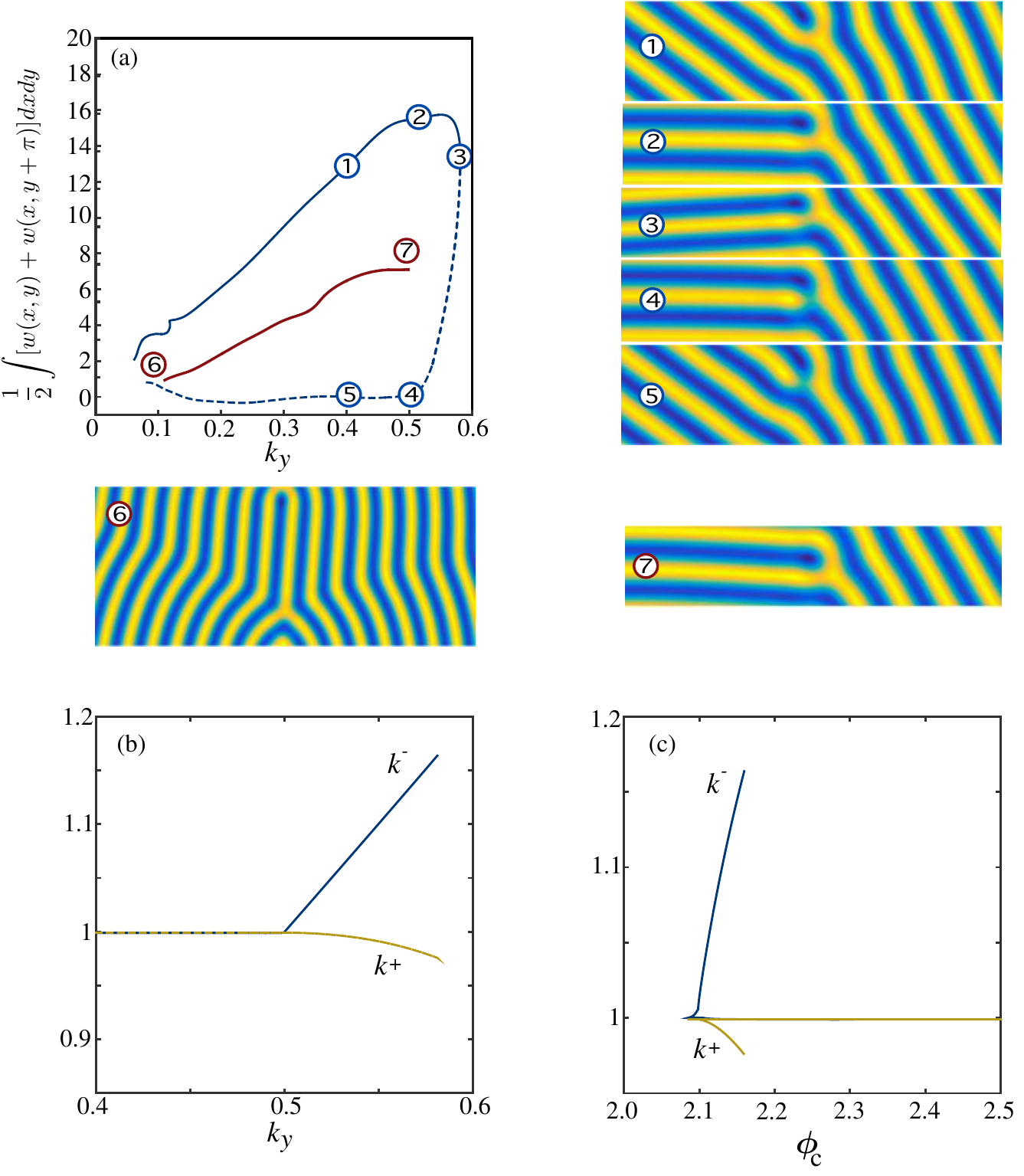}
	\caption{(a) Bifurcation diagram for $(2,1)$ (blue) and $(2,-1)$ (red) grain boundaries as a function of $k_y$ is shown. (b) and (c) show the same bifurcation diagram with $k^\pm$ versus $k_y$ or combined angle $\phi_\mathrm{c}$, respectively; for large angles, small $k_y$, $k^\pm=k_\mathrm{zz}$, which is also the wavenumber on the $(2,-1)$-branch. Sample plots correspond to labels on solutions branches. \label{f:21GB}}
\end{figure}

We focus on grain boundaries with $q_-\neq q_+$ and comment only briefly on other cases. Grain boundaries considered here break the reflection symmetry in $x$. Existence proofs are not known beyond the normal form approximation, which allows for a family of grain boundaries due to arbitrary relative phase shifts as described above. Our results strongly suggest existence of such grain boundaries for specific relative shifts and indicate some intriguing bifurcations. 

Practically, we compute grain boundaries asymptotic to striped patterns $u_\mathrm{s}(k_x^\pm x +q_\pm y;k^\pm)$ for given, fixed $q_\pm\in\Z$, with $k_y$ as our main bifurcation parameter.
Note that the distinction induced by the sign of $q_\pm$ is equivalent to a reflection in $x$. Including the parameter $k_y$, the family of stripes with $q_+$ is connected to the family with $-q_+$ through the horizontal stripes with $k_x=0$.

\paragraph{Dislocations and $(2,1)$ grain boundaries.}

\begin{figure}[htbp]
	\centering
	\includegraphics[width=0.65\linewidth]{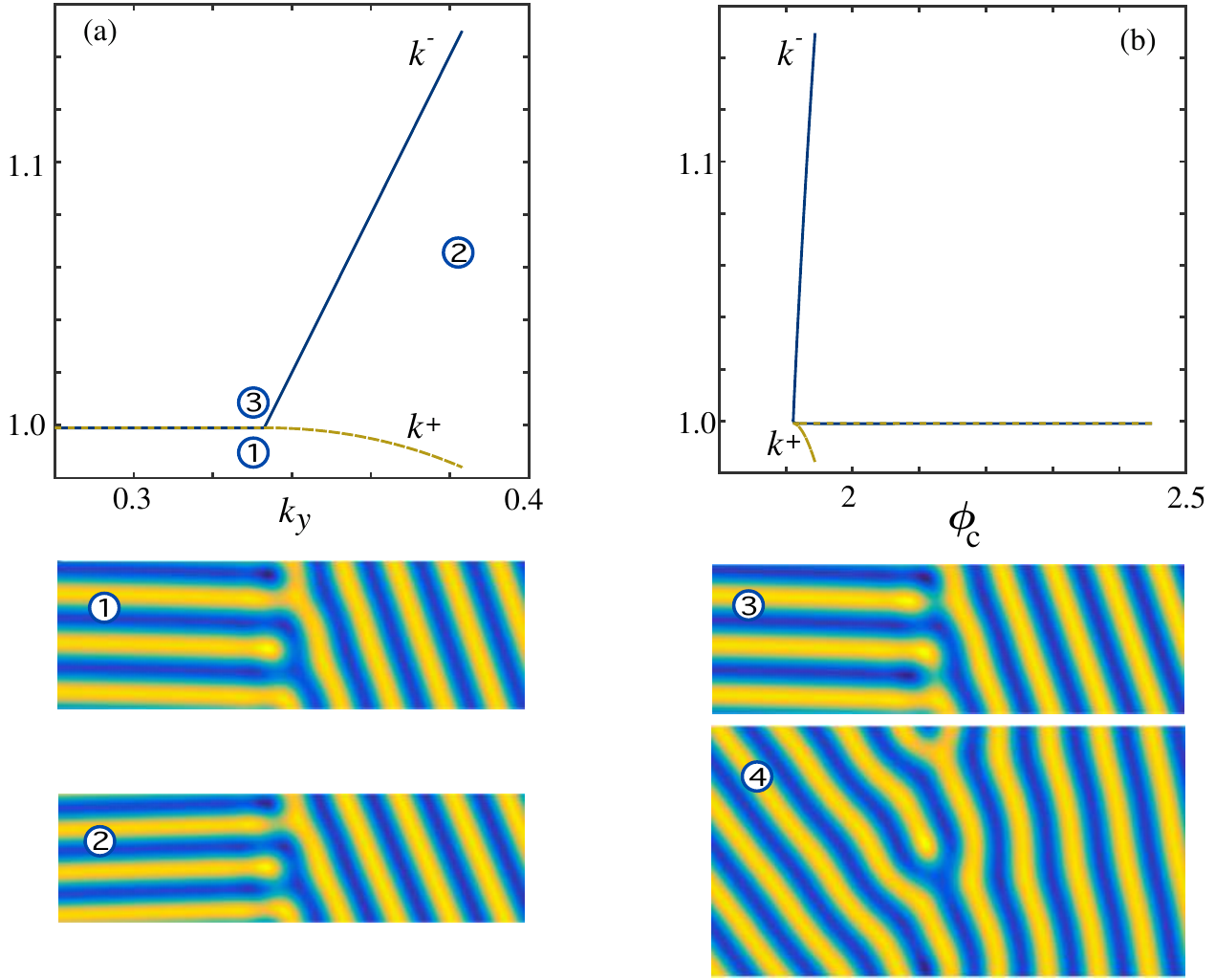}
	\caption{Selected wavenumbers and sample plots for $(3,1)$ grain boundaries. The bifurcation diagram is similar to Figure \ref{f:21GB} with a saddle-node bifurcation. Wavenumbers close to the saddle-node bifurcation \protect\raisebox{.5pt}{\textcircled{\protect\raisebox{-.9pt}{2}}} differ from the zigzag wavenumber, which is the selected wavenumber for stable and unstable branches away from the saddle-node, \protect\raisebox{.5pt}{\textcircled{\protect\raisebox{-.9pt}{1}}} and  \protect\raisebox{.5pt}{\textcircled{\protect\raisebox{-.9pt}{3}}}. The bottom sample plot \protect\raisebox{.5pt}{\textcircled{\protect\raisebox{-.9pt}{4}}} corresponds to $k_y=0.2$. \label{f:31GB}}
\end{figure}

We compute $(2,1)$ grain boundaries using the techniques introduced here. We continue in the parameter $k_y$ which forces an effective change in the relative angle when the wavenumber $k$ is kept fixed, for instance $k=k_\mathrm{zz}$. 

Figure~\ref{f:21GB} shows the bifurcation diagram in this case. For small values of $k_y$, the stripes are almost vertical and the slight discrepancy in angle is accommodated by a single dislocation (per vertical period) at the interface. Extending periodically, one sees that the grain boundary is in this way composed of evenly spaced dislocations. Increasing $k_y$ reduces the distance in this spacing and for values of $k_y$ closer to 0.5, the dislocations deform strongly. At $k_y\sim 0.5$, the marginally zigzag stable stripes are horizontal. Fixing $k=k_\mathrm{zz}$, they undergo a saddle-node bifurcation in $k_y$, where the slope of level lines changes sign. We do however \emph{not} see the grain boundaries following this saddle-node bifurcation. We see rather a saddle-node with an induced change of wavenumber in the far field, on both sides of the grain boundary. After the saddle-node, we see what appears to be a phase-mismatched $(2,1)$ grain boundary.

\paragraph{From $(2,1)$ to $(2,-1)$ grain boundaries.}

Plotting bifurcation diagrams against the combined angle as in Figure \ref{f:21GB} (c), one notices that the limiting case of a horizontal stripe on one side is an end point of a branch of grain boundaries. One can continue $(2,-1)$ grain boundaries in a similar fashion and finds again that the branch terminates at the horizontal grain boundaries. In fact, horizontal stripes with zigzag marginally stable wavenumber are also end points of $(1,-1)$ grain boundaries, in the limit of obtuse angle $\phi_c=\pi$. The zigzag instability corresponds to a Hamiltonian pitchfork bifurcation of the horizontal stripes in spatial dynamics \cite{haragus2007}, where heteroclinic orbits bifurcate. While we do not attempt here to analyze the heteroclinic bifurcation resulting from the interplay of this local pitchfork bifurcation and the global heteroclinic connection corresponding to the $(2,\pm1)$ grain boundary, our numerics clearly indicate that $(2,1)$- and $(2,-1)$-branches are not connected. Numerics are difficult since convergence rates in $L_x$ near the bifurcation point are slow (algebraic at $k_y=k_\mathrm{zz}/2$).

\paragraph{Pinning and $(0,1)$ grain boundaries.}

\begin{figure}[h!]
	\centering
	\includegraphics[width=0.7\linewidth]{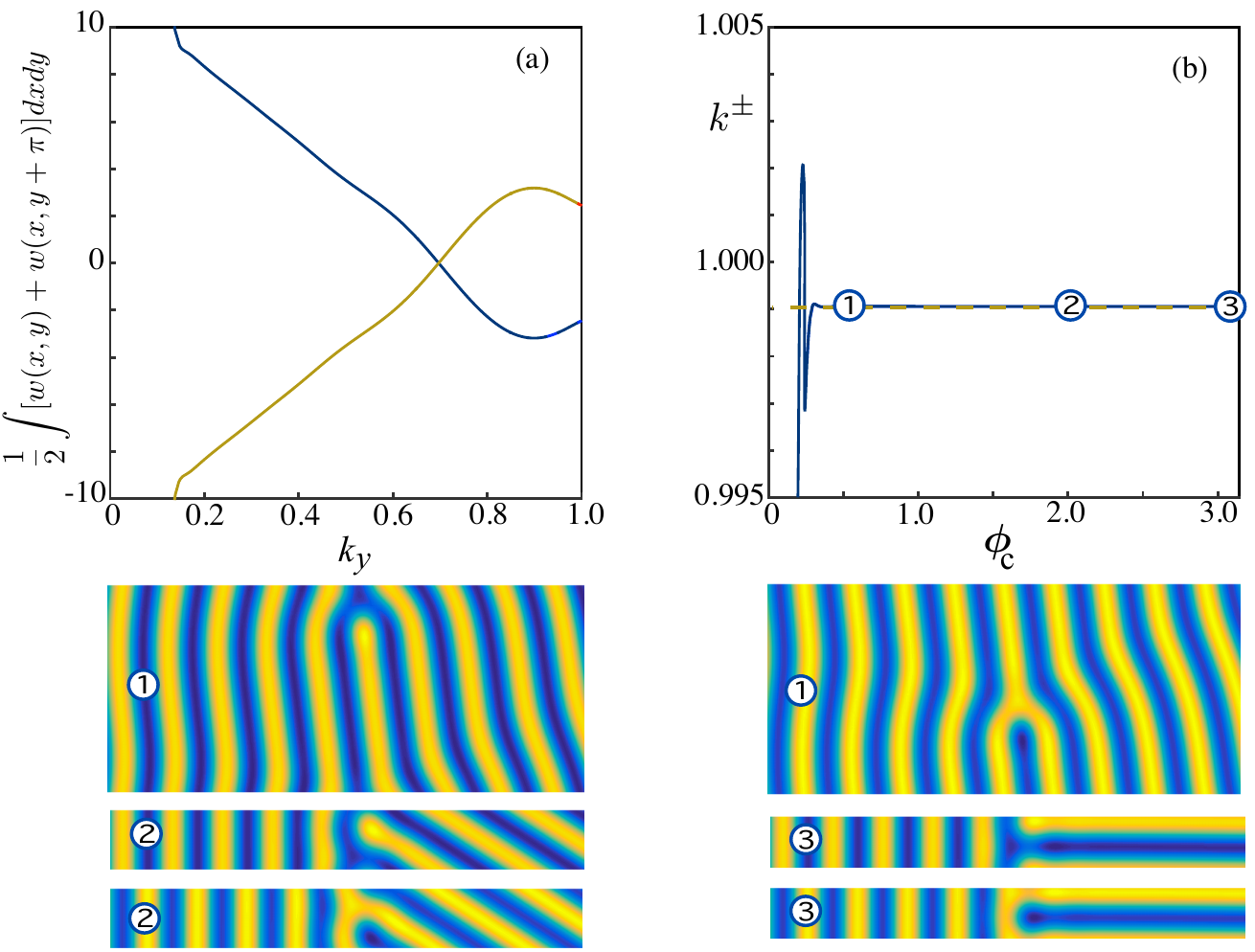}
	\caption{Continuation of vertical to slanted grain boundaries in $k_y$ (a) blue is with $\phi_-=0$ and gold is with $\phi_-=\pi$; selected wavenumber (b), energy (c), and sample plots of profiles (d). \label{f:vert_2}}
\end{figure}
Grain boundaries that are parallel to the stripes on one side of the grain boundary are somewhat special since a change of $k_y$ does not alter the orientation of the vertical stripes. We show results of our computations in Figure \ref{f:vert_2}. In particular, we did not detect any bifurcations; the selected wavenumber was the zigzag marginally stable wavenumber within numerical accuracy on both sides of the grain boundary. Interestingly, the energy is minimal for stripes perpendicular to the grain boundary. On the other hand, one finds a local minimum for small angles. Computations are somewhat delicate for both small angles and angles $\phi=\pi$ since the marginal zigzag stability induces slow decay towards stripes in these limiting cases. 

\paragraph{Grain boundaries involving $q>2$ --- preferred orientations.}

\begin{figure}[h]
	\centering
	\includegraphics[width=0.8\linewidth]{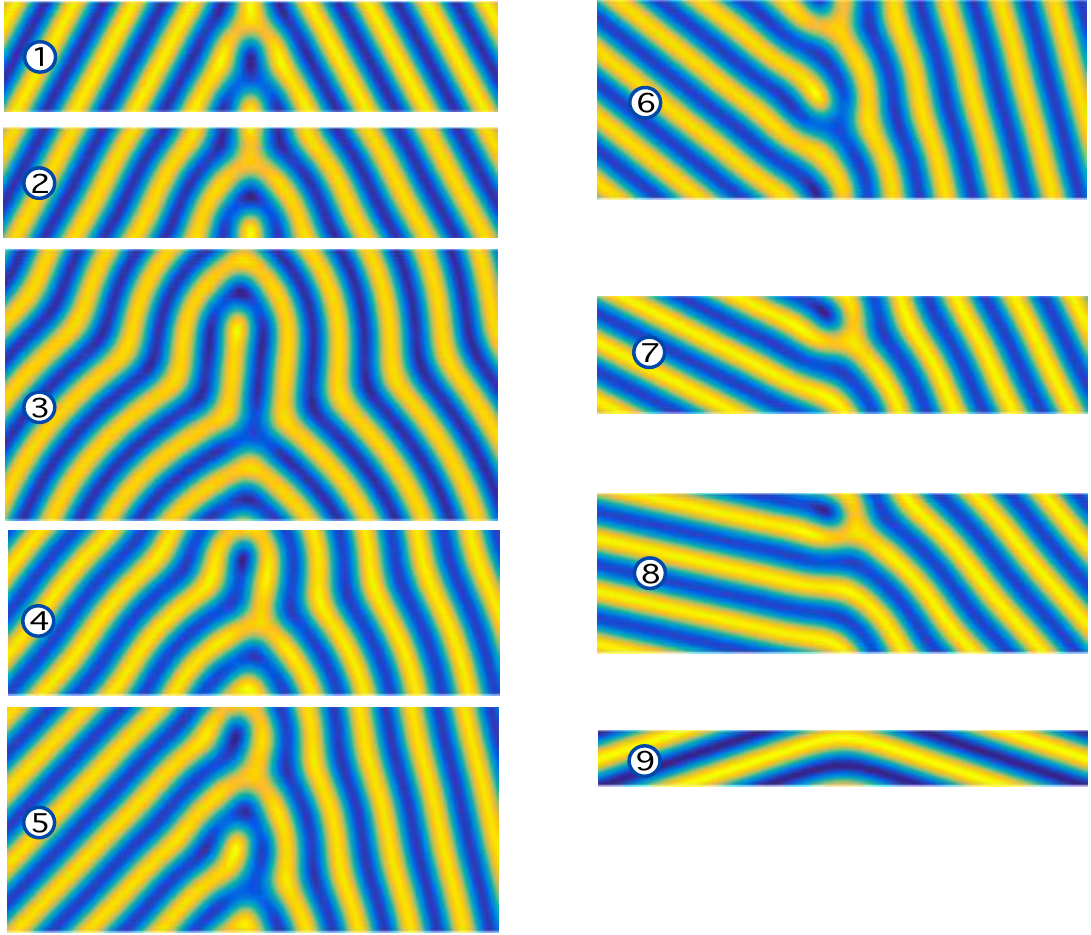}
	\caption{Asymmetric grain boundary profiles with decreasing energy 
 	\protect\raisebox{.5pt}{\textcircled{\protect\raisebox{-.9pt}{1}}} to \protect\raisebox{.5pt}{\textcircled{\protect\raisebox{-.9pt}{5}}} 
	for $q_->0$ and $q_+<0$ and $\phi_\mathrm{c}=1$, and decreasing energy  
	\protect\raisebox{.5pt}{\textcircled{\protect\raisebox{-.9pt} {6}}} to \protect\raisebox{.5pt}{\textcircled{\protect\raisebox{-.9pt} {9}}} $q_-,q_+>0$ with $\phi_\mathrm{c}=2.5$.
	\protect\raisebox{.5pt}{\textcircled{\protect\raisebox{-.9pt} {1}}} $(q_-,q_+)=(1,-1),k_y=0.48$,
	\protect\raisebox{.5pt}{\textcircled{\protect\raisebox{-.9pt} {2}}} $(q_-,q_+)=(1,-1),k_y=0.48$ on the pitchfork branch,
	\protect\raisebox{.5pt}{\textcircled{\protect\raisebox{-.9pt} {3}}} $(q_-,q_+)=(3,-2),k_y=0.19$,
	\protect\raisebox{.5pt}{\textcircled{\protect\raisebox{-.9pt} {4}}} $(q_-,q_+)=(2,-1),k_y=0.31$,
	\protect\raisebox{.5pt}{\textcircled{\protect\raisebox{-.9pt} {5}}} $(q_-,q_+)=(3,-1),k_y=0.23$,
	\protect\raisebox{.5pt}{\textcircled{\protect\raisebox{-.9pt} {6}}} $(q_-,q_+)=(3, 1),k_y=0.26$,
	\protect\raisebox{.5pt}{\textcircled{\protect\raisebox{-.9pt} {7}}} $(q_-,q_+)=(2, 1),k_y=0.33$,
	\protect\raisebox{.5pt}{\textcircled{\protect\raisebox{-.9pt} {8}}} $(q_-,q_+)=(3, 2),k_y=0.33$,
	\protect\raisebox{.5pt}{\textcircled{\protect\raisebox{-.9pt} {9}}} $(q_-,q_+)=(1,-1),k_y=0.95$.
	\label{f:asym_GBs}}
\end{figure}

\begin{figure}[h]
	\centering
	\includegraphics[width=0.45\linewidth]{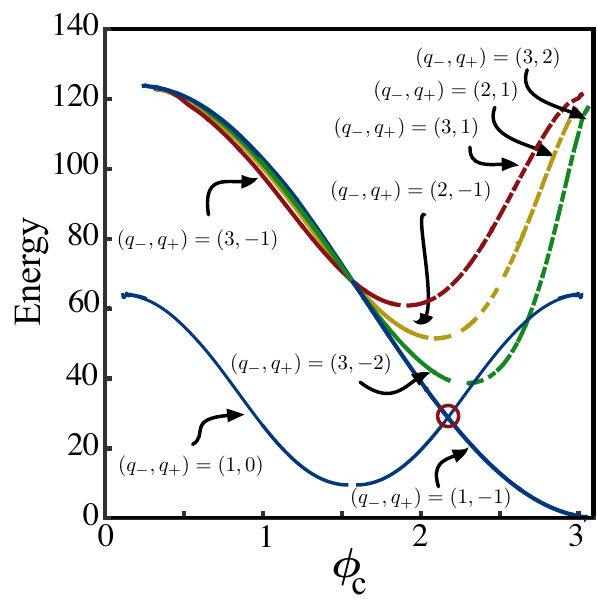}
	\caption{Energy of grain boundaries depending on the combined angle $\phi_\mathrm{c}$, for several values of $(q_-,q_+)$.  Note that the simplest symmetric $(1,-1)$ grain boundary is energetically preferred for angles $\phi_\mathrm{c}>\pi/2$. Intersection of energy curves $(0,1)$ and $(1,-1)$ occurs at $\phi_c=2.16$.}\label{f:gbe2}
\end{figure}

From a continuation and bifurcation point of view, $(3,1)$ grain boundaries behave in a completely analogous fashion as $(2,1)$ grain boundaries, as illustrated in Figure \ref{f:31GB}. The actual grain boundaries are now composed of two dislocations, conjugate by the parity $u\mapsto -u$, as becomes most apparent in the limit of small $k_y$. We computed more generally $(q_-,q_+)$ grain boundaries and show sample profiles in  Figure~\ref{f:asym_GBs}.

In order to determine preferred orientations of grain boundaries, one defines and fixes a combined  angle $\phi_\mathrm{c}$,
\[
\phi_\mathrm{c}=\left\{\begin{array}{ll}
\pi - (\phi_++\phi_-), & q_->0>q_+,\\
\phi_-+\phi_+, & q_-,q_+>0;
\end{array}\right.
\]
see also Figure \ref{f:gbsc}. Varying now $q_\pm$ while fixing $\phi_\mathrm{c}$, one attempts to find an orientation of the grain boundary that minimizes the energy per unit interfacial length. Grain boundaries in Figure~\ref{f:asym_GBs} are displayed with decreasing energy, top to bottom. Figure \ref{f:gbe2} shows the energy of grain boundaries depending on the combined angle, for several choices of $(q_-,q_+)$.

We separated the asymmetric grain boundaries into two groups: those with $q_-$ and $q_+$ of opposite and those with $q_-$ and $q_+$ of same sign. In each of the columns in Figure~\ref{f:asym_GBs}, we order the grain boundaries in decreasing order of renormalized energy~(\ref{e:energy_SHr}) for the same combined angle $\phi_\mathrm{c}=1$ and $\phi_\mathrm{c}=2.5$, respectively. It is interesting to note that for the grain boundaries with $q_-$ and $q_+$ of opposite sign, the symmetric grain boundary $(q_-,q_+)=(1,-1)$ is not always the most energetically preferred. 
For acute angles, the energy of grain boundaries appears to decrease with the ratio $q_-/q_+$, indicating a tendency of stripes on one side of the boundary to align with the interface. The preferred orientation is then actually the vertical $(1,0)$ grain boundary between slanted and vertical stripes. For obtuse angles, small ratios $q_-/q_+$ appear to be preferred, with the defect-free weakly bent $(1,-1)$ grain boundary having significantly less energy than other grain boundaries. We found a critical angle of $\phi_c^*=2.16$, such that $(1,0)$ grain boundaries are preferred for $\phi_c<\phi_c^*$ and $(1,-1)$ grain boundaries for $\phi_c>\phi_c^*$. 

\paragraph{Grain boundaries --- lists from stacking defects.}

\begin{figure}[h!]
	\centering
	\includegraphics[width=0.8\linewidth]{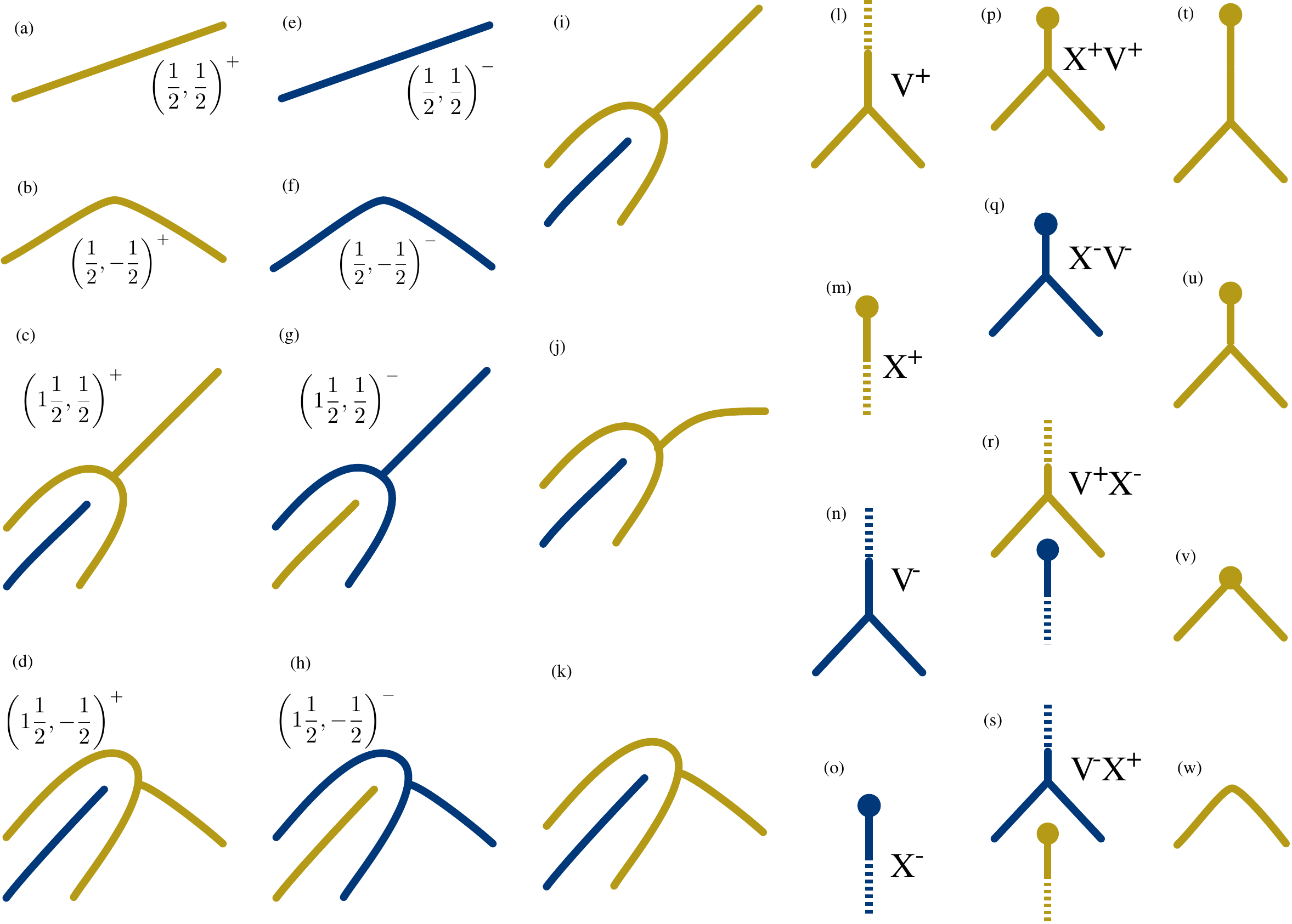}
	\caption{Top Schematic plot of building blocks for grain boundaries in (a)--(h) for asymmetric grain boundaries and (l)--(o) for symmetric grain boundaries. Transitions between building blocks (i)--(k) and (t)--(w), and bound states (p)--(s); see explanations in text.
	}\label{f:schematic}
\end{figure}

\begin{figure}[h!]
	\centering
	\includegraphics[width=0.8\linewidth]{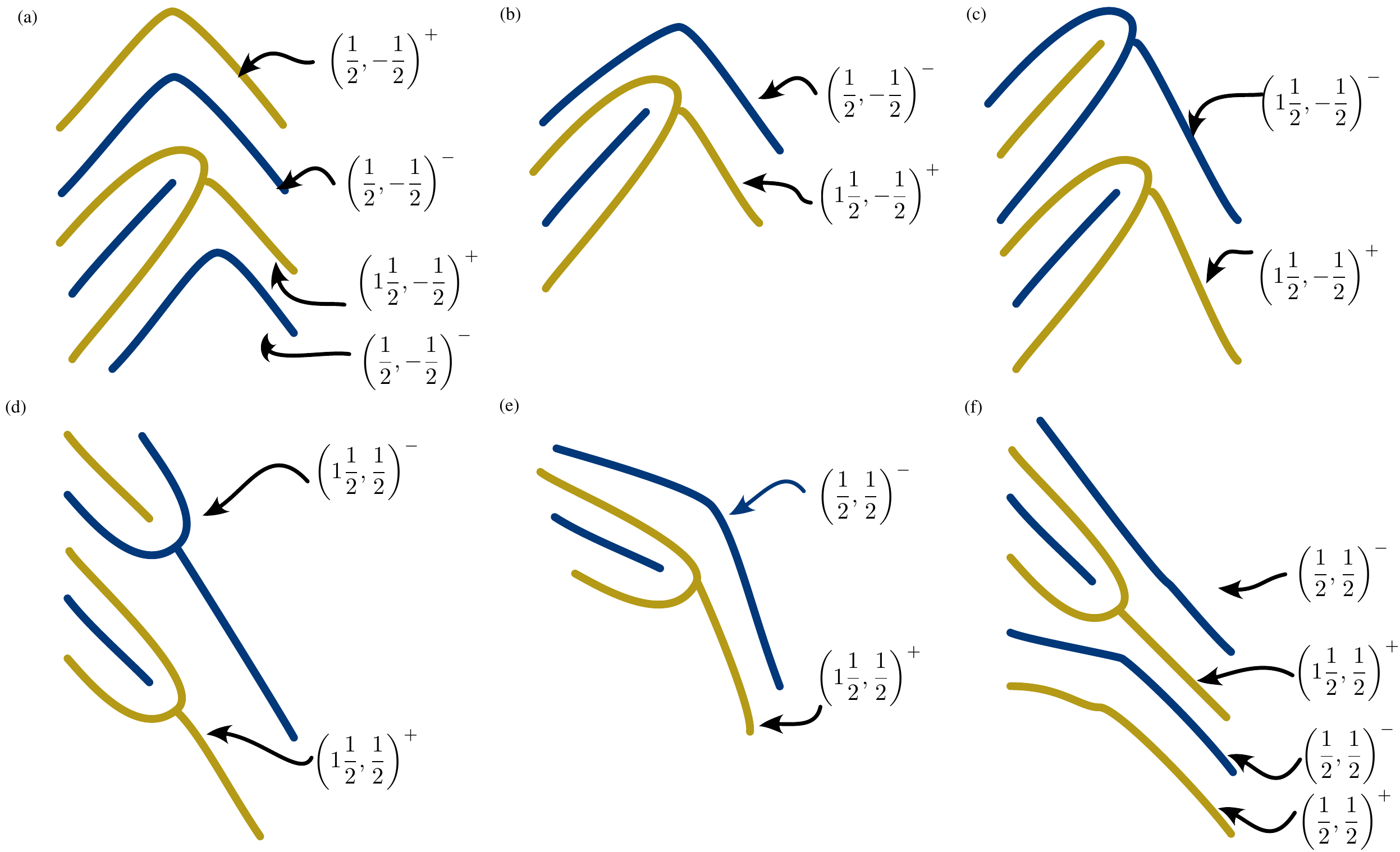}
	\caption{Building asymmetric grain boundaries from Figure \ref{f:asym_GBs} from the building blocks in Figure \ref{f:schematic}. Grain boundaries \protect\raisebox{.5pt}{\textcircled{\protect\raisebox{-.9pt} {3}}}--\protect\raisebox{.5pt}{\textcircled{\protect\raisebox{-.9pt} {5}}}, top row, and \protect\raisebox{.5pt}{\textcircled{\protect\raisebox{-.9pt} {6}}}--\protect\raisebox{.5pt}{\textcircled{\protect\raisebox{-.9pt} {8}}}, bottom row, of type $(3,-2),\,(2,-1),\,(3,-1)$ and $(3,1),\,(2,1),\,(3,2)$, respectively. 
	}\label{f:stack}
\end{figure}
One clearly notices in Figure \ref{f:asym_GBs} that, particularly for acute angles, grain boundaries can be interpreted as composed of defects such as dislocations, or convex and concave disclinations. In Figure \ref{f:schematic}, we list the basic building blocks. Because of the parity symmetry $u\mapsto -u$, all building blocks come in two versions, $+$ and $-$. The left column shows, beyond the simple ``bend'' $(\frac{1}{2},-\frac{1}{2})^+$ (half a vertical period of an obtuse $(1,-1)$ grain boundary), dislocations and bent dislocations as bound states of disclinations; the second column shows the parity-conjugates. The third column shows a continuous deformation between straight and bent dislocations, as would arise in a transition from $(2,-1)$ to $(2,1)$ grain boundaries. Note however that the bifurcation diagrams in Figures \ref{f:21GB} and \ref{f:31GB} show that these transitions do not actually occur along a branch of grain boundaries. The fourth column shows the elementary disclinations, concave (V) and convex (X) in both parities. The last two columns show bound states and transitions, most notably dislocations as convex-concave bound state $V^+X^-$ and $V^-X^+$, and the annihilation of a convex-concave disclination pair as observed along the parity-shift symmetry broken branch of $(1,-2)$ grain boundaries.

\begin{figure}[h]
	\centering
	\includegraphics[width=0.95\linewidth]{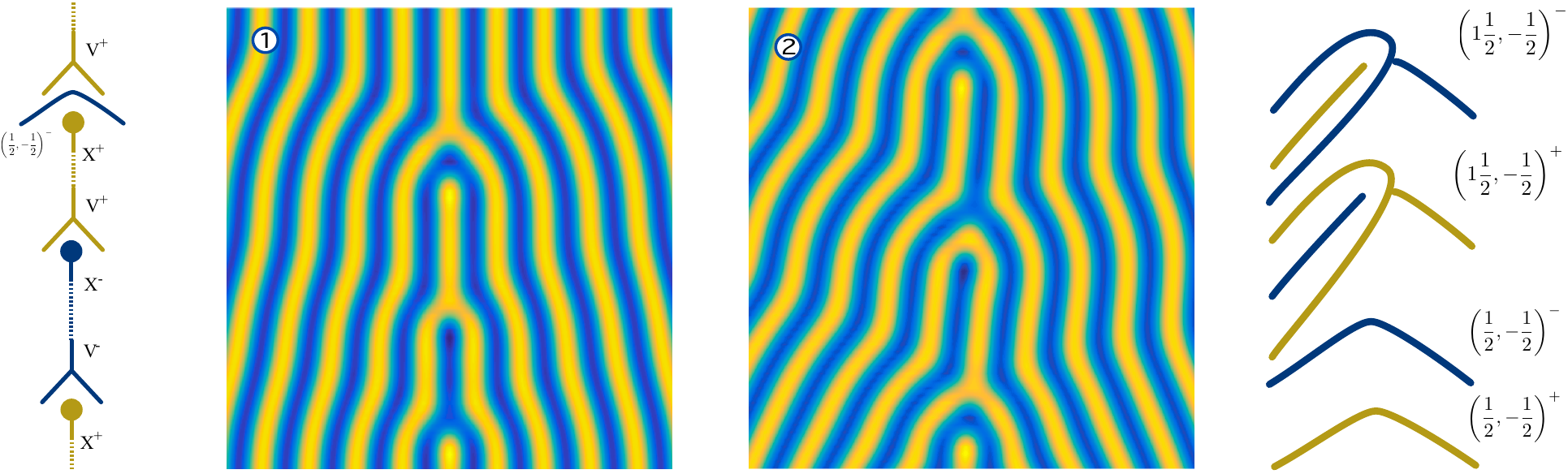}
	\caption{Anomalous grain boundaries of type $(2,-2)$ and $(4,-2)$ which are \emph{not} simply doubled grain boundaries of type $(1,-1)$ and $(2,-1)$, respectively, obtained from direct simulations of the initial-value problem in a doubly periodic box; $k_y=0.1$, $x\in (-10\pi,10\pi)$, shown is $x\in [-30,30]$.
	}\label{f:4222}
\end{figure}

Using the building blocks from Figure \ref{f:schematic}, we can in fact construct formally all grain boundaries computed in Figure \ref{f:asym_GBs} through simple ``stacking''; see  Figure \ref{f:stack}.
Similarly, one can construct acute $(1,-1)$ grain boundaries as $V^+X^-|V^-X^+$ stacks and the parity-shift broken branch as $V^+|(\frac{1}{2},-\frac{1}{2})^-|X^+$. 

One can now easily predict a variety of ``new'' grain boundaries, such as a $(4,-2)$-grain boundary obtained from stacking $(1\frac{1}{2},-\frac{1}{2})^+| (1\frac{1}{2},-\frac{1}{2})^-|(\frac{1}{2},-\frac{1}{2})^+| (\frac{1}{2},-\frac{1}{2})^-$,    or a $(2,-2)$-grain boundary obtained by stacking   $V^+|(\frac{1}{2},-\frac{1}{2})^-|X^+|V^+X^-|V^-X^+$. Such grain boundaries can indeed be observed as stable interfaces as demonstrated in Figure \ref{f:4222}.

\paragraph{Energetically preferred grain boundaries.}
Given our results above, one can anticipate energetically preferred shapes of grain boundaries for a given angle. For acute angle grain boundaries, combined angle $\phi_\mathrm{c}<\pi/2$,  Figure \ref{f:gbe2} suggests that grain boundaries where stripes on one side are parallel to the interface are energetically preferred. Otherwise, weak symmetric bending $(1,-1)$ grain boundary appear to be preferred. One can rationalize this effect by observing the defects generated at the boundary. The $(1,0)$ grain boundaries can be composed of an $X^-|V^+$ sequence in each period, whereas a $(3,-1)$ grain boundary, say, is composed of a $(1\frac{1}{2},\frac{1}{2})^+|(1\frac{1}{2},\frac{1}{2})^-$ sequence which in turn consists of 4 disclinations $V^+|X^-|V^-|X^+$, suggesting that the larger number of defects leads to a higher interfacial energy. We emphasize however that a simple count of ``defects/unit length'' will generally not give a sharp criterion. For instance, the energy of grain boundaries per unit length increases as the combined angle becomes more acute, although the vertical period increases and hence the number of defects per unit length decreases. One factor here certainly is the fact that disclinations are strongly deformed from their ideal shape as an isolated point defect when angles are acute. 

Given two grain orientations and an interface orientation, $\phi_\mathrm{c}<\pi/2$, it is then conceivable that grain boundaries are built from piecewise straight grain boundaries that are parallel to either left or right stripes, interspacing both orientations of the grain boundary such that the resulting prescribed angle is achieved. We suspect that pinning effects in the interaction between defects that build the grain boundary will prevent coarsening of these piecewise straight segments of grain boundaries. 

For small differences in the grain orientation, obtuse angles, our results confirm the suspicion that defect-free bending is the energetically preferred mode of accommodating the orientation mismatch. 

Finally, Figure \ref{f:vert_2} suggests that out of all the grain boundaries with stripes parallel to the interface, grain boundaries with angle $\pi/2$ are preferred. It is however not clear how, starting with random patches of grain orientations, configurations with only such grain boundaries could emerge.

\subsection{Other grain boundaries}\label{s:other}

We think of the examples shown here as the ``simplest'' grain boundaries for given angles $\phi_\pm$. As we noticed when ``stacking'' defects, grain boundaries can often be thought of as composed of simpler defects such that interaction dynamics are in equilibrium. It appears that these interaction dynamics allow for a multitude of pinning effects, and our goal here is to elucidate some more obvious examples of this. 

\begin{figure}[h!]
	\centering
	\includegraphics[width=0.6\textwidth]{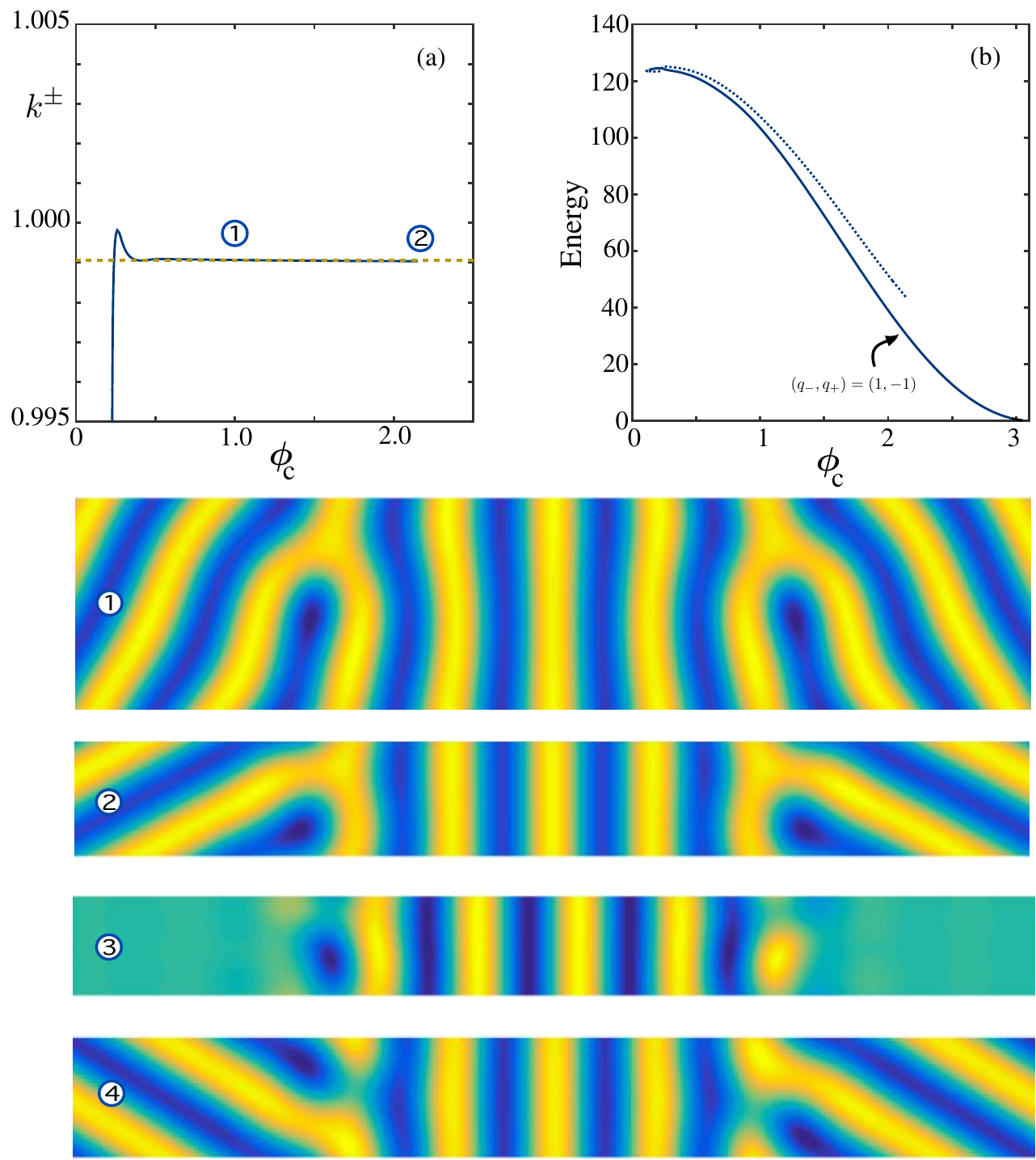}
	\caption{Selected wavenumber and energy of symmetric grain boundaries with a core of vertical stripes (top). Sample plots shown below \protect\raisebox{.5pt}{\textcircled{\protect\raisebox{-.9pt} {1}}}--\protect\raisebox{.5pt}{\textcircled{\protect\raisebox{-.9pt} {2}}}. The branch appears to terminate at $\phi_\mathrm{c}\sim 2.14$ at a bifurcation point (eigenfunction shown in \protect\raisebox{.5pt}{\textcircled{\protect\raisebox{-.9pt} {3}}}). Also shown \protect\raisebox{.5pt}{\textcircled{\protect\raisebox{-.9pt} {4}}})  a sample plot of a $(1,1)$ ``grain boundary'' with vertical core.
	}\label{f:vcore}
\end{figure}

\paragraph{Symmetric grain boundaries with vertical stripe core.}

In the simplest case of weak bending, when grain boundaries can be described through phase modulation equations near the zigzag instability, the Cahn-Hilliard equation, stationary profiles include, in addition to the ``heteroclinic'' grain boundaries also homoclinic orbits, which can be viewed as concatenations of two heteroclinic orbits. Those ``step''-like double-knees are unstable in the Cahn-Hilliard modulation approximation due to possible coarsening. More importantly, they do not connect the energy-minimizing marginally zigzag stable stripes. 

In Figure \ref{f:vcore}, we show $(1,-1)$ grain boundaries that contain a core of vertical stripes. They can be thought of as concatenations of $(1,0)$ and $(0,-1)$ grain boundaries, respectively. It is interesting to notice that again the asymptotic wavenumber does not appear to depend on the angle and is, within numerical accuracy, the zigzag marginally stable one. The energy is however larger than the energy of the simpler $(1,-1)$ grain boundaries that we computed before. Curiously these grain boundaries do not appear to continue to the weak bending regime. We also note that one might expect the interaction between grain boundaries to be exponentially decaying in the distance, such that any deviations from the zigzag wavenumber caused by the interaction could be beyond numerical resolution. We also computed $(1,1)$ grain boundaries resulting from a concatenation of $(1,0)$ and $(0,1)$ grain boundaries, resulting in a homoclinic type solution, with equal grain orientation on on both sides. 
\begin{figure}[h!]
	\centering
	\includegraphics[width=0.6\textwidth]{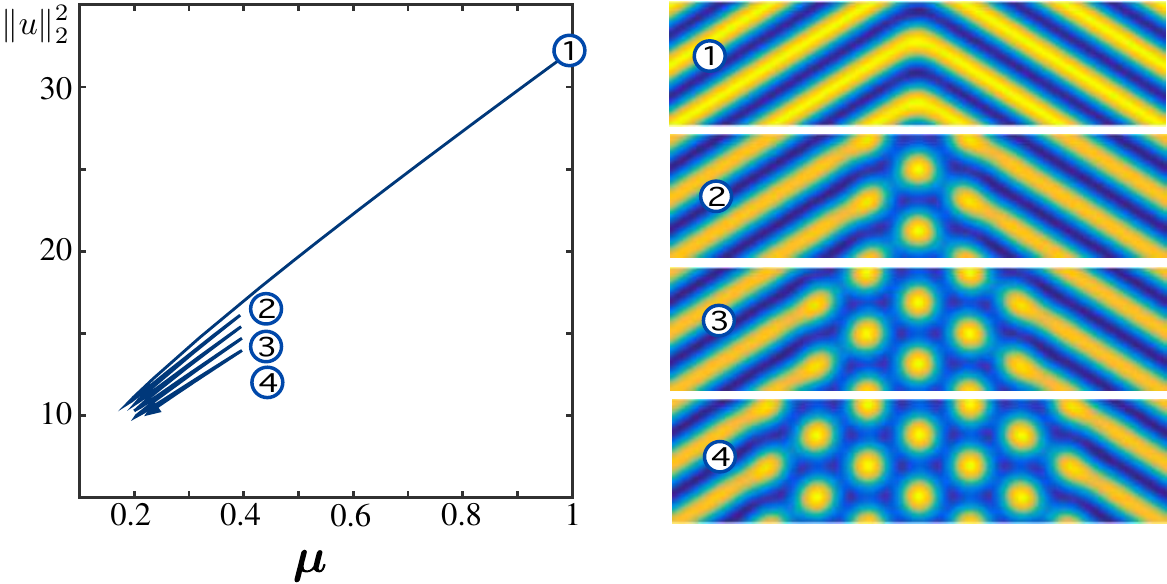}
	\caption{Snaking symmetric grain boundaries for a modified quadratic-cubic nonlinearity, $f(u) = u^2 - u^3$; $k_y=0.85$.}\label{f:hex}
\end{figure}

\paragraph{Snaking grain boundaries.}

One would expect stronger pinning effects when hexagonal spot patterns are involved. We confirm this in a continuation computation involving a quadratic-cubic nonlinearity, which allows for hexagonal patterns. The results of a sample computation are shown in Figure \ref{f:hex}. For small $\mu$, hexagons nucleate at the tip of the ``knee'', where stripes are most unstable, and cause a saddle-node bifurcation. We expect there to be a plethora of grain boundaries for small $\mu$. We note that grain boundaries involving hexagonal patterns had also been discussed in \cite{malomed1990}. Since analysis there was performed within the amplitude equation framework, snaking and pinning aspects, in particular for moderate values of $\mu$ are not analyzed there. 

\section{Discussion}\label{s:6}

We presented a robust framework for the computation of grain boundaries from a path-following perspective and explored grain boundaries in the Swift-Hohenberg equation within this framework. The path-following perspective shows that complexity of grain boundaries increases for acute angles, with the creation of defects and pinning effects in their interaction. For obtuse angles, weak, defect-free, reflection-symmetric  bending is the energetically preferred interface structure. For acute angles, disclinations and dislocations form at interfaces and create a wealth of structures. Asymmetric  grain boundaries with stripes parallel to the grain boundary on one side appear to be energetically preferred in this case and we anticipate zigzag patterns in the shape of the actual grain boundary interface for minimum energy interfaces. While we outline a system for cataloging such grain boundaries, we did not attempt an exhaustive description. In all regimes, our analysis raises a number of interesting questions, both from an analytic and a computational perspective. 

\paragraph{Stability.} 
Our stability analysis here is somewhat rudimentary, relying on a simple eigenvalue computation in the bounded domain. Since the essential spectrum touched the imaginary axis, it would be more appropriate to track eigenvalues using some variant of the Evans function and its extension into the essential spectrum, allowing for a precise tracking of eigenvalues near the origin and how they turn into resonance poles upon crossing into the essential spectrum.  We expect that such a computation could be accomplished using  a decomposition quite analogous to our far-field-core decomposition here; see for instance the analysis in  \cite{poganscheel,fayescheel,ssmorse}. Analytically, one may wish to start studying stability of small-amplitude grain boundaries, at least up to the neutral eigenvalue corresponding to phase matching and non-adiabatic effects beyond the normal form, possibly first on the spectral level, preparing for a nonlinear stability proof. 

Since the assumption on $y$-periodicity is technical, one would also want to study stability with $y\in\R$. On the spectral level, this would add a Bloch-wave parameter $\sigma$ accounting for $y$-modulations. Even theoretically, it is not clear how zeros of the Evans function would behave for long-wavelength modulations, $\sigma\sim 0$, in particular, if one can associate a bending stiffness $d$, $\lambda\sim -d\sigma^2$ to grain boundaries.

\paragraph{Bifurcations.}
Our computations point to a number of interesting bifurcations. We mention here in particular bifurcations involving horizontal stripes, which, due to the marginal zigzag instability are neutrally stable with a length 4 Jordan block at the origin, bifurcations at the core such as the parity-shift breaking and the snaking bifurcations, as well as gluing bifurcations that produce grain boundaries with striped cores as in Figure \ref{f:vcore}. It would also be interesting to explore the effect of non-variational terms, such as the shift of the wavenumber away from marginal zigzag stability. Phenomenologically, it would be interesting to find parameter regions where grain boundaries select zigzag unstable stripes, thus inducing cascades of bending analogous to the cascades of spiral waves generated by far-field breakup instabilities \cite{ssbreakup}.

\paragraph{Point defects.}
In light of the complexity of bifurcation diagrams for $(p,q)$ grain boundaries when $p,q$ are not necessarily small, and for acute angles, one would wish to describe grain boundaries in terms of point defects and their interaction properties. One therefore would like to implement analogous far-field-core decompositions for point defects. Preliminary theoretical results in this direction have been obtained in \cite{jaramillo13,jsw16}, providing in particular an implicit function theorem near stripes in the presence of localized inhomogeneities and hinting at systematic multipole approximations for the phase in the far-field. 

\paragraph{Beyond stripes and Swift-Hohenberg.}
While our computations address only one specific parameter value in one specific equation, one could hope for some wider-ranging implications. For this, it would be interesting to perform more extensive comparisons with modifications of Swift-Hohenberg, including parity-breaking and non-variational terms. Beyond the prototypical Swift-Hohenberg equation, one could investigate systems with striped phases other than Swift-Hohenberg, such as polymer, phase separation,  or reaction-diffusion systems  \cite{gbpolymer,phasefield,turinggb}. 

On the other hand, grain boundaries have been extensively studied in nonlinear elasticity material models, although commonly not from the point of view taken here, using path following and idealizations to infinite domains, and resolving the fine crystalline structure. As pointed out above, one would expect pinning effects to be more complex in hexagonal lattices, leading to complicated snaking phase diagrams; see for example~\cite{sutton1995} for a general exposition on the interfaces in crystalline matter. 

In this context, but also in the Swift-Hohenberg example, it would be interesting to study the response of grain boundaries to inhomogeneities and external forces. Localized inhomogeneities or impurities could be readily incorporated into our framework as $(x,y)$-dependent forcing terms, which would break translational symmetry. As a result, one would not expect Goldstone modes in the kernel and cokernel, changing the parameter counts from Section \ref{s:1}. Even in the absence of grain boundaries, inhomogeneities induce phase shifts and, in some cases, wavenumber shifts \cite{jsw16}. One might expect such effects also for grain boundaries, at least in the weak bending regime of $(1,-1)$ grain boundaries. Similarly, external forces could be modelled as boundary conditions at finite $x=\pm L$, that select wavenumbers different from the energy minimizing $k_\mathrm{zz}$ \cite{morrissey2015}. In the absence of grain boundaries, incompatible imposed strains induce a drift of stripes in the direction of the gradient of the strain, as can be seen in the scalar phase diffusion equation 
\[
\Theta_t=\Theta_{xx},\ x\in(-L,L),\qquad  \Theta_x\big|_{x=\pm L}=\pm 1,\qquad \Theta(t,x)=(2L)^{-1}\left(x^2 -2t\right).
\]
One may expect such an induced drift for grain boundaries, at least in the weak bending regime, since grain boundaries themselves can be thought of as strain (or wavenumber) selecting, imposing thus an effective,  possibly incompatible Neumann boundary condition at $x=0$.

%
%

\section{Appendix}

We describe in more technical detail the far-field-core decomposition that is key to our numerical continuation procedure. We start by assuming that there exists a grain boundary $u_*$ with asymptotic stripes with wavenumbers $k^\pm$. 

\subsection{The linearization at a grain boundary --- Fredholm properties}

Linearizing (\ref{e:gbr}) at a grain boundary, we obtain the elliptic operator
\[
\mathcal{L}_* u=-(\partial_x^2+k_y^2\partial_y^2+1)^2 u + \mu u - 3u_*^2u, \qquad (x,y)\in \R\times (0,2\pi),
\]
equipped with periodic boundary conditions in $y$. We typically think of $\mathcal{L}_*$ as a closed, densely defined operator on $L^2$, with domain $H^4$. Using Weyl sequence arguments, one readily finds that $\mathcal{L}_*$ is not Fredholm due to the presence of non-localized elements of the kernel, $\partial_x u_*$ and $\partial_y u_*$. 

The presence of continuous spectrum of $\mathcal{L}_*$ at $\lambda=0$ reflects the numerically observed slow, diffusive adjustment of stripes to perturbations. Focusing on the simple coherent structure rather than the plethora of dynamics nearby, we choose exponential weights that allow us to directly separate far-field behavior from the core of the grain boundary. Consider therefore exponentially weighted spaces $H^k_\eta(\R\times (0,2\pi))$, with norm
\[
\|u(x,y)\|_{H^k_\eta}=\|u(x,y)\rme^{\eta \langle x\rangle}\|_{H^k},
\]
where $\langle x\rangle=\sqrt{x^2+1}$.  

We will see next that, typically, $\mathcal{L}_*$ is Fredholm on $L^2_\eta$ for $\eta>0$, sufficiently small. Therefore, define the asymptotic operators $\mathcal{L}_\pm$, where $u_*$ in the definition of $\mathcal{L}_*$ is replaced by $u_\pm(x,y):=u_\mathrm{s}^\pm(k_x^\pm x+k_y^\pm y;k^\pm)$. Also, define the exponentially weighted spaces $H^k_{\eta,>}$ via 
\[
\|u(x,y)\|_{H^k_{\eta,>}}=\|u(x,y)\rme^{\eta x}\|_{H^k}.
\]

\begin{Proposition}
The operator $\mathcal{L}_*$ is Fredholm on $L^2_\eta$ if and only if $\mathcal{L}_\pm$ are invertible in $L^2_{\pm\eta,>}$.
\end{Proposition}
\begin{Proof}
The proof is a direct application of the closed range lemma; see for instance \cite{robbin1995}.
\end{Proof}
In order to better understand the operators $\mathcal{L}_\pm$, we introduce Bloch waves. Consider
\[
\hat{\mathcal{L}}_\pm(\nu)=-(\partial_y^2+(\partial_x+\nu)^2+1)^2+\mu-3u_\mathrm{s}^\pm,
\]
with periodic boundary conditions on $(0,2\pi/k_x)\times (0,2\pi/k_y)$. Classical Bloch wave theory \cite{reed1980} says that 
\[
\mathrm{spec}_{L^2_{\eta,>}}\,(\mathcal{L}_\pm)=\bigcup_{\nu\in -\eta+\rmi[0,k_x)}\mathrm{spec}_{L^2_\mathrm{per}}\,(\hat{\mathcal{L}}_\pm(\nu)).
\]
In order to understand the spectrum of $\hat{\mathcal{L}}_\pm(\nu)$, we relate to the linearization at a ``straight'', non-rotated stripe, $u_\mathrm{s}(k^\pm x)$.
Consider therefore the Floquet-Bloch linearization
\[
\hat{\mathcal{L}}(\nu_x,\nu_y)=-(\nu_y^2+(\partial_x+\nu_x)^2+1)^2+\mu-3u_\mathrm{s}^\pm(k x;k).
\]
Within the stability region $k\in(k_\mathrm{zz},k_\mathrm{eck})$, and for $\Re\nu_x,\Re\nu_y$ small, the spectrum of $\hat{\mathcal{L}}(\nu_x,\nu_y)$ is strictly negative, bounded away from the origin, except for a simple eigenvalue $\lambda$ close to the origin when $\eta_x,\eta_y\sim 0$, with expansions
\[
\lambda(\nu_x,\nu_y)=d_\parallel \nu_x^2+d_\perp \nu_y^2+\rmO(|\nu_x|^4+|\nu_y|^4),
\]
with positive constants $d_\parallel$ and $d_\perp$. At $k=k_\mathrm{zz}$, we find 
\[
\lambda(\nu_x,\nu_y)=d_\parallel \nu_x^2-d_\perp \nu_y^4+\rmO(|\nu_x|^4+|\nu_y|^6);
\]
see for instance \cite{cross1993,mielke1997}.

\begin{Lemma}
For $k\in [k_\mathrm{zz},k_\mathrm{eck}]$ and arbitrary $k_x<k$, the operators $\mathcal{L}_\pm$ are invertible in $L^2_{\pm\eta,>}$ for $\eta>0$, sufficiently small. 
\end{Lemma}
\begin{Proof}
Using Fourier-Bloch decomposition, 
\[
u(x,y)=\sum_{m\in\Z} u_m(k_x x + k_y y;k)\rme^{\rmi m k_y y},
\]
the operator $\mathcal{L}_\pm$ diagonalizes over $m$, with diagonal entries 
\[
\hat{\mathcal{L}}_\pm(\nu)=-\left((k_y\partial_\xi+\rmi m k_y)^2+(k_x\partial_\xi+\nu)^2+1\right)^2+\mu-3u_\mathrm{s}^\pm,
\]
with $2\pi$-periodic boundary conditions in $\xi$. These operators are equal to $\hat{\mathcal{L}}(\nu_x,\nu_y)$ when choosing 
\begin{align*}
\nu_x&=(\nu k_x + \rmi m k_y^2)/k,\\
\nu_y^2&=\nu^2-m^2k_y^2-\nu_x^2,
\end{align*}
where, of course, $\nu_x,\nu_y$ may be complex. We see that for $\nu$ small and $m\neq 0$, $\nu_x$ is not small so that the critical eigenvalue $\lambda(\nu_x,\nu_y)$ does not vanish. For $m=0$, 
\[
\nu_x^2=\nu^2\frac{k_x^2}{k_x^2+k_y^2},\qquad 
\nu_y^2=\nu^2\frac{k_y^2}{k_x^2+k_y^2},
\]
so that 
\begin{equation}\label{e:qu}
\lambda\sim \nu^2
\end{equation}
for small $\nu$ within the stability region. 
\end{Proof}
We remark that the assumptions in the lemma are not sharp. Invertibility follows when the rotated stripes are marginally stable with $y$-periodic bouundary conditions. We next proceed to determine the Fredholm index of the linearization. 
\begin{Lemma}
For $\eta>0$ sufficiently small, the Fredholm index of the linearization at a grain boundary  is $-2$ in $L^2_\eta$ and it is $+2$ in $L^2_{-\eta}$,  provided that the asymptotic stripes are marginally stable ($k_x\neq 0$ for $k=k_\mathrm{zz}$, and $k_y\neq 0$ for $k=k_\mathrm{eck}$).
\end{Lemma}
\begin{Proof}
Suppose first that $\eta>0$. The Fredholm index can be computed by counting the signed crossings of multipliers through the origin during a homotopy from $\hat{\mathcal{L}}_+(\nu)$ to $\hat{\mathcal{L}}_-(-\nu)$. From the preceding Lemma, we can homotope between $\hat{\mathcal{L}}_+(\nu)$ and $\hat{\mathcal{L}}_-(\nu)$ without crossings. Homotoping from $\nu$ to $-\nu$, we see precisely the double zero multiplier from \eqref{e:qu} cross the origin, which readily gives the desired result on the Fredholm index. Since the linearization $\mathcal{L}_*$ is self-adjoint in $L^2$, $\mathcal{L}_*$ is Fredholm of index 2 in $L^2_{-\eta}$ for $\eta>0$, small. 
\end{Proof}
Similar results have been proven in \cite{sandstede2004}, where spatial dynamics rather than spectral flow arguments were employed. Also, the discussion there centers around the case of stripes with non-vanishing group velocities, when the dispersion relation \eqref{e:qu} contains a linear term in $\nu$ and crossings are simple. The present case is most similar to the case of contact defects discussed there.

\subsection{Transverse grain boundaries}
The translational modes  $\partial_xu_*$ and $\partial_y u_*$ span a two-dimensional subspace of the kernel of $\mathcal{L}_*$ in $L^2_{-\eta}$, $\eta>0$ (note that both are linearly independent since otherwise the grain boundary would be a one-dimensional pattern, consisting of a simple stripe). On the other hand, since asymptotic wave vectors $\underline{k}^\pm$ are different, we cannot find a linear combination of  $\partial_xu_*$ and $\partial_y u_*$ that is exponentially localized. 
\begin{Hypothesis}[Transverse GB]\label{h:tgb}
Assume that the kernel of  $\mathcal{L}_*$ in $L^2_{-\eta}$, $\eta>0$ sufficiently small, is two-dimensional.
\end{Hypothesis}
Note that, as a consequence, the kernel of $\mathcal{L}_*$ in $L^2_\eta$, $\eta>0$, is trivial, and the cokernel is spanned by  $\partial_xu_*$ and $\partial_y u_*$. 

We emphasize that the grain boundaries found in the truncated normal form near onset, $\mu\sim 0$, are not transverse. The additional normal form symmetry leads to an element in the kernel of $\mathcal{L}_*$ in $L^2_{\eta}$. In our numerical computations, here, we found that this hypothesis is typically satisfied as one might expect. 

\subsection{Far-field matching and robustness}
Since the linearization in exponentially weighted spaces is Fredholm, we would like to employ the implicit function theorem in order to continue grain boundaries in parameters. For negative weights, the linearization is onto but the nonlinearity is not defined. For positive weights, the negative index indicates that additional free variables are necessary to solve. These are naturally given through wavenumbers and phase shifts in the far field. We exploit those with an ansatz
\[
u(x,y)=w(x,y)+\chi_+(x)u_\mathrm{s}^+(x,y)+\chi_-(x)u_\mathrm{s}^-(x,y),\quad u_\mathrm{s}^\pm(x,y)=
u_\mathrm{s}(k^\pm_xx+q_\pm y+\varphi^\pm;k^\pm),
\]
where, the smooth cut-off functions $\chi_\pm$ satisfy 
\[
\chi_\pm(x)=1,\ \pm x>d+1,\qquad 
\chi_\pm(x)=0,\ \pm x<d,
\]
for some (arbitrary) $d>0$. Here,  $w\in H^4_\eta$, $k^\pm_x$, and $\varphi_\pm$ are free variables. The asymptotic wavenumbers $k^\pm$ satisfy
\[
(k^\pm)^2=(k_x^\pm)^2+(q_\pm k_y)^2,
\]
where $k_y$ is a free parameter and $q_\pm\in\Z$ are fixed integers. 

Substituting this ansatz into the stationary Swift-Hohenberg equation gives
\begin{equation}\label{e:shm0}
\mathcal{L}\left(w+\sum_\pm \chi_\pm u_\mathrm{s}^\pm\right)-\left(w+\sum_\pm \chi_\pm u_\mathrm{s}^\pm\right)^3=0,
\end{equation}
or, equivalently, after subtracting the equations for $u_\mathrm{s}^\pm$,
\begin{equation}\label{e:shm}
\mathcal{L}w+\sum_\pm\left[\mathcal{L},\chi_\pm\right]u_\mathrm{s}^\pm 
- \left[\left(w+\sum_\pm \chi_\pm u_\mathrm{s}^\pm\right)^3-\left(\sum_\pm \chi_\pm u_\mathrm{s}^\pm\right)^3\right]
+\left[\sum_\pm\chi_\pm \left(u_\mathrm{s}^\pm\right)^3-\left(\sum_\pm\chi_\pm u_\pm\right)^3  \right]=0,
\end{equation}
where
\[
\mathcal{L}=-(\partial_x^2+k_y^2\partial_y^2+1)^2+\mu,\quad   \mbox{and}\; \ \ [A,B]u=A(Bu)-B(Au).
\]
We consider the left-hand side of \eqref{e:shm} as a (locally defined) nonlinear operator 
\[
F_w:H^4_\eta \times \R^6\to L^2_\eta, \quad (w,k_x^-,k_x^+,k_y,\varphi^-,\varphi^+)\to F_w(w,k_x^-,k_x^+,k_y,\varphi^-,\varphi^+).
\]
Note that $F_w$ is well defined since terms not involving $w$ are given by commutators between cut-off and differential operators and nonlinearities, hence compactly supported, as one can easily check by setting $\chi_+=1,\chi_-=0$ in \eqref{e:shm}. 

Moreover, $F_w$ is readily seen to be a smooth function and the derivative with respect to $w$ at a grain boundary $u_*=w+\sum_\pm\chi_\pm u_\mathrm{s}^\pm$ is the linearization we discussed before,
\[
\partial_wF_w=\mathcal{L}_*,
\]
so that $DF_w=(\partial_wF_w,\partial_{k_x^\pm,k_y,\varphi^\pm}F_w)$ is Fredholm of index 3 by Fredholm bordering theory. 
\begin{Lemma}\label{l:onto}
The linearization $D_{w,k_x^\pm}F_w$ at a transverse grain boundary is invertible. The derivatives $D_{\varphi^\pm}F_w$ belong to the range of $D_wF_w$.
\end{Lemma}
\begin{Proof}
First notice that solutions of $F_w=0$ come in families induced by translations in $x,y$,
\begin{align*}
u_*(x+\tau_x,y+\tau_y)&=\tilde{w}(x,y)+\sum_\pm\chi_\pm(x)u_\mathrm{s}^\pm(k^\pm_x x + k^\pm_y y + \tilde{\varphi}^\pm),\\
\tilde{w}(x,y)&=w(x+\tau_x,y+\tau_y)+\sum_\pm \left(\chi_\pm(x+\tau_x)-\chi_\pm(x)\right)u_\mathrm{s}^\pm(k^\pm_x x + k^\pm_y y+\tilde{\varphi}^\pm),\\
\tilde{\varphi}^\pm&=\varphi^\pm+k_x^\pm\tau_x+k_y^\pm\tau_y.
\end{align*}
Differentiating with respect to $\tau_{x/y}$ gives 
\[
(w,k^-,k^+,\varphi^-,\varphi^+)=(\partial_x w+\sum_\pm\chi_\pm'u_\mathrm{s}^\pm,0,0,k_x^+,k_x^-),\qquad 
(w,k^-,k^+,\varphi^-,\varphi^+)=(\partial_y w,0,0,k_y^+,k_y^-).
\]
As a consequence, using that $(k_x^+,k_y^+)$ and $(k_x^-,k_y^-)$ are linearly independent, we find that $\partial_{\varphi^\pm}F_w\in\Rg\partial_wF_w$.

We proceed to show that $\partial_{k_x^\pm}F_w$ span the cokernel of $\partial_wF_w$. Suppose this was not the case. Then there would exist $\alpha_\pm\in\R$, so that $\sum_\pm \alpha_\pm\partial_{k_x^\pm}F_w=\mathcal{L}_*w_0$, for some $w_0\in H^4_\eta$. Since $\partial_{k_x^\pm}F_w=\mathcal{L}_*w_k^\pm$, with $w_k^\pm=\chi_\pm\frac{\rmd}{\rmd k_x^\pm}u_\mathrm{s}^\pm$, we conclude that 
$\mathcal{L}_*(w_0+\sum_\pm\alpha_\pm w_k^\pm)=0$. However, $\sum_\pm\alpha_\pm w_k^\pm\in H^4_{-\eta}$, with linear growth, and  $\sum_\pm\alpha_\pm w_k^\pm\not\in H^4_{\eta}$ for $(\alpha_+,\alpha_-)\neq 0$, since the supports of $w_k^\pm$ are disjoint, so that $w_0+\sum_\pm\alpha_\pm w_k^\pm\neq 0$. We have thus found an element in the kernel of $\mathcal{L}_*$ in $H^4_{-\eta}$, in contradiction to the transversality assumption.
\end{Proof}

As an immediate consequence, we can use the implicit function theorem to conclude robustness of grain boundaries.
\begin{Corollary}\label{c:t}
Transverse grain boundaries come in families, parameterized by $k_y$. Such families persist under small perturbations of the equation, such as variations of the parameter $\mu$. 
\end{Corollary}
The choice of $k_y$ as parameter and $k_x^\pm$ as variables is somewhat arbitrary. Modifying Lemma \ref{l:onto}, one could also choose combinations such as $k_y,k_x^+$ as variables. 
%
%
\begin{Remark}[Symmetric grain boundaries]\label{r:s}
Analytic existence results are only available for symmetric grain boundaries, $u_*(x,y)=u_*(-x,y)$, or $u_*(x,y)=-u_*(-x,y)$, where of course $q_\pm=\pm 1$. In this case, one can restrict to even (or odd) functions, $H^4_\mathrm{even}$ and finds that $\mathcal{L}_*$ is Fredholm of index $\pm 1$  in $H^4_{\mp \eta}$. Using $k_x^+=k_x^-=:k_x$ as an additional variable, we can then continue transverse grain boundaries in $k_y$. Alternatively, one can write $k_x=mk_y$ and consider the grain orientation $m$ as parameter and the wavenumber $k=\sqrt{1+m^2}k_y$ as variable. Again, following the proof of Lemma \ref{l:onto}, one finds that the linearization is onto so that transverse grain boundaries can be characterized by a selected wavenumber as a function of the grain orientation $k=k(m)$.
\end{Remark}


\subsection{Approximating grain boundaries in finite intervals --- theory}
\label{s:3}

The previous considerations promote a view of an isolated grain boundary as a coherent structure in an idealized infinite system. In particular, one would like to compute grain boundaries and transmission relations such as $(k_x^+,k_x^-)$ as functions of $k_y$ as in Corollary \ref{c:t}, or selected wavenumbers $k(m)$ as a function of grain orientation as in Remark \ref{r:s}.  Our point of view is to assume the existence of a grain boundary and compute with error bounds in finite domains $\Omega_{L_x}=(x,y)\in (-L_x,L_x)\times (0,2\pi)$. We will therefore construct a suitable approximation to $F_w$ in such finite domains. 

Consider therefore
\begin{align}\label{e:shmL}
F_w^{L_x}&(w,k_x^\pm,k_y,\varphi^\pm)\nonumber\\
=&\mathcal{L}w+\sum_\pm\left[\mathcal{L},\chi_\pm\right]u_\mathrm{s}^\pm 
- \left[\left(w+\sum_\pm \chi_\pm u_\mathrm{s}^\pm\right)^3-\left(\sum_\pm \chi_\pm u_\mathrm{s}^\pm\right)^3\right]
+\left[\sum_\pm\chi_\pm \left(u_\mathrm{s}^\pm\right)^3-\left(\sum_\pm\chi_\pm u_\mathrm{s}^\pm\right)^3  \right],
\end{align}
on $w\in H^4(\Omega_{L_x})$, with periodic boundary conditions in $y$ and boundary conditions at $x=\pm L_x$ to be specified later. With standard elliptic boundary conditions, say Dirichlet $w=w_{xx}=0$ at $x=\pm L_x$, the linearization $\partial_wF_w^{L_x}$ is a Fredholm operator of index 0 by standard elliptic regularity. This linear operator is, however, very ill-conditioned, with norms for the inverse growing as $L^2$ at best; see for instance \cite{ssabs} and our discussion in Section \ref{s:1}. The view point introduced in the preceding section will prove more effective, also in setting of finite but large domains. 

Consider therefore boundary conditions $\mathcal{B}_\pm(w,w_x,w_{xx},w_{xxx})=0$ at $x=\pm L_x$, for all $y\in [0,2\pi)$, such that $-\Delta^2$ is Fredholm of index 0 when equipped with periodic boundary conditions in $y$ and $\mathcal{B}_\pm$. Examples are Dirichlet, Neumann, or mixed boundary conditions for the Laplacian, or clamped ($w=w_x=0$) boundary conditions. In addition, consider phase conditions
\begin{equation}\label{e:phc}
p_\pm w=\int_{x_\pm}^{x_\pm+2\pi/k_x^\pm} \psi_\pm(x)w(x)\rmd x,
\end{equation}
with suitable choice of $\psi_\pm$ and $x_\pm$. Then, by simple Fredholm bordering and perturbation theory, the linear operator $(\partial_{w,k_x^\pm}F_w^{L_x},p_\pm w):H^4_\mathrm{bc}\times \R^2\to L^2\times \R^2$ is Fredholm of index 0. The next hypothesis is necessary for stability and convergence of the decomposition.

\begin{Hypothesis}[Transverse boundary conditions]\label{h:bc}
We assume that the linearization at the asymptotic stripe pattern $\mathcal{L}_+$ , posed on $(-\infty,x_*)\times (0,2\pi)$, equipped with boundary conditions $\mathcal{B}_+$ at $x=x_*$ and phase condition $p_+$ does not possess a kernel in $H^4_\eta$ for any $x_*\in[0,2\pi/k_x^+)$. We also require that $\mathcal{L}_-$ , posed on $(x_*,\infty)\times (0,2\pi)$, equipped with boundary conditions $\mathcal{B}_-$ and phase condition $p_-$ does not possess a kernel in $H^4_\eta$ for any $x_*\in[0,2\pi/k_x^-)$.
\end{Hypothesis}

These assumptions, Hypotheses \ref{h:tgb} and \ref{h:bc}, put us in a situation analogous to \cite[Proposition 2.11]{morrissey2015}, where bifurcation diagrams in bounded domains were predicted up to exponentially small errors in the domain size $L_x$. Following the strategy of proof there should therefore yield exponential convergence of the solution $(w,k^\pm)$ of the truncated boundary-value problem \eqref{e:shmL} to the actual grain boundary
\[
\left|w_{L_x}-w_*\right|_{H^4}+\left|k_{x,L_x}^\pm-k_{x,*}^\pm\right|=\rmO\left(\rme^{-\delta L_x}\right).
\]
Note that in the language of \cite{morrissey2015}, grain boundaries are purely wavenumber selecting, yielding trivial strain-displacement relations $k=k_*$, $\varphi\in S^1$, and the additional phase condition imposed here selects one representative from the family of admissible solutions.

Lack of transversality manifests itself in the appearance of boundary layers, that is, $w$ ceases to be uniformly exponentially localized as $L_x\to\infty$. A similar nonlinear statement gives exponential bounds on the truncation error. 

In practice, we will choose Dirichlet boundary conditions for $w$ and phase conditions with $\psi_\pm=(u_\mathrm{s}^\pm)'$, effectively eliminating the incorporation of phase shifts and wavenumber corrections into the corrector $w$. Numerical observations described in the paper indicate that this choice does indeed yield transverse boundary conditions. We did notice failure of transversality when choosing Neumann boundary conditions, for acute angles between stripes.

\bibliographystyle{abbrv}
\bibliography{grain_boundaries_bib}

\end{document}